\documentclass[10pt]{article}
\usepackage{multicol}
\usepackage{graphicx}
\usepackage{amsmath}

\begin{document}

\begin{center}
{\Large\bf Extracting kinetic freeze-out temperature and radial
flow velocity from an improved Tsallis distribution}

\vskip0.5cm

Hai-Ling Lao$^{a}$, Fu-Hu Liu$^{a,}${\footnote{E-mail:
fuhuliu@163.com; fuhuliu@sxu.edu.cn}}, and Roy A. Lacey$^{b}$

{\small\it $^a$Institute of Theoretical Physics, Shanxi
University, Taiyuan, Shanxi 030006, China

$^b$Departments of Chemistry \& Physics, Stony Brook University,
Stony Brook, NY 11794, USA}
\end{center}

\vskip0.25cm

{\bf Abstract:} We analyze the transverse momentum ($p_T$) spectra
of identified particles ($\pi^{\pm}$, $K^{\pm}$, $p$, and $\bar
p$) produced in gold-gold (Au-Au) and lead-lead (Pb-Pb) collisions
over a $\sqrt{s_{NN}}$ (center-of-mass energy per nucleon pair)
range from 14.5 GeV [one of the Relativistic Heavy Ion Collider
(RHIC) energies] to 2.76 TeV [one of the Large Hadron Collider
(LHC) energies]. For the spectra with a narrow $p_T$ range, an
improved Tsallis distribution which is in fact the Tsallis
distribution with radial flow is used. For the spectra with a wide
$p_T$ range, a superposition of the improved Tsallis distribution
and an inverse power-law is used. Both the extracted kinetic
freeze-out temperature ($T_0$) and radial flow velocity
($\beta_T$) increase with the increase of $\sqrt{s_{NN}}$, which
indicates a higher excitation and larger expansion of the
interesting system at the LHC. Both the values of $T_0$ and
$\beta_T$ in central collisions are slightly larger than those in
peripheral collisions, and they are independent of isospin and
slightly dependent on mass.
\\



\vskip1.0cm

\begin{multicols}{2}

{\section{Introduction}}

Transverse momentum ($p_T$) spectra of identified particles
produced in proton-proton, proton-nucleus, and nucleus-nucleus
collisions at high energies are important quantities measured in
experiments. In particular, gold-gold (Au-Au) collisions at the
Relativistic Heavy Ion Collider (RHIC) and lead-lead (Pb-Pb)
collisions at the Large Hadron Collider (LHC) and other high
energy nucleus-nucleus collisions have been providing us excellent
chances to study the signals and features of quark-gluon plasma
(QGP), the properties of multi-particle production, and the
characteristics of the interesting system. In the study of $p_T$
spectra, we can obtain some useful information which contains, but
is not limited to, the effective temperatures ($T$) when the
interacting system emit different particles, chemical freeze-out
temperature ($T_{ch}$) based on particle ratios, kinetic
freeze-out temperature ($T_0$), and transverse or radial flow
velocity ($\beta_T$).

On the extraction of $T$, one can use different functions or
distribution laws such as the standard (Boltzmann, Fermi-Dirac, or
Bose-Einstein) distribution [1--4], the Tsallis distribution
[4--10], and others. On the extraction of $T_{ch}$, one can use
the particle ratios due to different normalization constants for
different identified particle spectra in a given $p_T$ range. On
the extractions of $T_0$ and $\beta_T$, one can use the
distributions which contains simultaneously $T_0$ and $\beta_T$
such as the blast-wave model [11, 12] and an improved Tsallis
distribution (the Tsallis distribution with radial flow) [13, 14],
as well as an alternative method which is used in our recent works
[15--17] and partly used in the previous literature [11, 18--20].

The blast-wave model [11, 12] is a traditional and current method
which has wide applications. This model makes the simple
assumption that particles are locally thermalized in a hard-sphere
uniform density source at a kinetic freeze-out temperature and are
moving with a common collective transverse radial flow velocity
field. The improved Tsallis distribution is a new method which is
suggested by Sahoo and his colleagues [13, 14] and has a few
applications. This model is based on the Tsallis distribution and
introduces radial flow in it. Thus, the temperature parameter in
the improved Tsallis distribution is the kinetic freeze-out
temperature $T_0$. We are curious to use the improved Tsallis
distribution in the present work to obtain some distinctive
conclusions. The alternative method is partly a new one, in which
the extraction of $T_0$ has some applications [11, 18--20]. In the
alternative method, $T_0$ is regarded as the intercept in the
linear relation between $T$ and $m_0$, and $\beta_T$ is regarded
as the slope in the linear relation between $\langle p_T \rangle$
and $\langle m \rangle$, where $m_0$, $\langle m \rangle$, and
$\langle p_T \rangle$ denote the rest mass, mean moving mass, and
mean $p_T$, respectively.

We are interested in the consistency and differences of the three
methods in their results. In fact, their differences are larger
than their consistency. The blast-wave model [11, 12] and the
improved Tsallis distribution [13, 14] have their assumptions and
pictures respectively. The alternative method itself is
independent of models, though its result $T_0$ depends on $T$
which depends on distribution laws of $p_T$, and its result
$\beta_T$ is independent of distributions due to the same $\langle
p_T \rangle$ and $\langle m \rangle$ for given experimental
spectra. Generally, the Tsallis distribution results in lower $T$
and $T_0$ than the standard distribution. Also, the blast-wave
model [11, 12] and the improved Tsallis distribution [13, 14]
result in lower $T_0$ than the alternative method when using the
standard distribution. Different distributions are in fact
different `thermometers' or `thermometric scales' and
`speedometers'.

In this paper, we shall use the improved Tsallis distribution [13,
14] to fit $p_T$ spectra of identified particles ($\pi^{\pm}$,
$K^{\pm}$, $p$, and $\bar p$) produced in Au-Au collisions at the
RHIC and Pb-Pb collisions at the LHC. The center-of-mass energy
per nucleon pair, $\sqrt{s_{NN}}$, considered by us is from 14.5
GeV to 2.76 TeV. After fitting the experimental data measured by
the STAR [21, 22], PHENIX [20, 23--28], and ALICE Collaborations
[29], we analyze the tendency of parameters.

The rest part of this paper is structured as follows. A brief
description of the formalism is presented in section 2. Results on
comparisons with experimental data and discussion are given in
section 3. Finally, we summarize our main observations and
conclusions in section 4.
\\

{\section{The formalism}}

High energy collisions are a complex process in which many
emission sources are formed. The sources with the same excitation
degree may form a local equilibrium state which can be described
by the standard distribution. For different equilibrium states
which have different excitation degrees, different temperature
parameters may be used. Generally, a two- or three-component
standard distribution can describe the $p_T$ spectrum in a not too
wide $p_T$ range, which reflects the temperature fluctuation of
the interacting system. At the same time, a two- or
three-component standard distribution can be described by the
Tsallis distribution with the parameters which contains mainly $T$
and the entropy index $q$.

The Tsallis distribution, i.e. the $p_T$ distribution in the
Tsallis statistics, has more than one forms. In a recent work
[10], five forms of Tsallis and related distributions are
collected. We have the Tsallis distribution at mid-rapidity
($y\approx 0$) as follows
\begin{equation}
f_1(p_T)=C_1 p_T m_T \bigg[1+(q-1)\frac{m_T}{T}\bigg]^{-q/(q-1)},
\end{equation}
where $C_1$ is the normalization constant which results in
$\int_0^{\infty} f_1(p_T)dp_T=1$, $m_T=\sqrt{p_T^2+m_0^2}$ is the
transverse mass, and $m_0$ is the rest mass. The chemical
potential is not included in Eq. (1) due to its small effect on
the $p_T$ distribution. Other four Tsallis-related distributions
at mid-rapidity are
\begin{equation}
f_2(p_T)=C_2 p_T m_T \bigg[1+(q-1)\frac{m_T}{T}\bigg]^{-1/(q-1)},
\end{equation}
\begin{equation}
f_3(p_T)=C_3 p_T \bigg[1+(q-1)\frac{m_T}{T}\bigg]^{-q/(q-1)},
\end{equation}
\begin{equation}
f_4(p_T)=C_4 p_T \bigg[1+(q-1)\frac{m_T}{T}\bigg]^{-1/(q-1)},
\end{equation}
and
\begin{equation}
f_5(p_T)=C_5 p_T \bigg[ 1+\frac{q-1}{T} (m_T-m_0)
\bigg]^{-1/(q-1)},
\end{equation}
where $C_{2,3,4,5}$ denote the different normalized constants
which result respectively in $\int_0^{\infty}
f_{2,3,4,5}(p_T)dp_T=1$ for different distribution forms which are
rearranged comparing with ref. [10]. Although we have used the
same symbols, the values of $T$ (or $q$) in Eqs. (1)--(5) are
different from each other.

The above Tsallis-related distributions can be used to describe
the $p_T$ spectra of particles produced in soft excitation process
which occurs between gluons and/or sea quarks and contributes to a
not too wide $p_T$ range. Because of their similarity, one of them
is enough for the description of soft process. From the similarity
and self-consistency to the standard distribution, Eqs. (1) and
(3) are the favorable choices. However, the values of $T$ obtained
from the Tsallis-related distributions are only effective
temperatures which contain the contributions of thermal motion and
flow effect together. To disentangle the thermal motion and flow
effect, an alternative method can be used in the case of analyzing
$p_T$ spectra of identified particles.

Fortunately, Sahoo and his colleagues [13, 14] have introduced the
radial flow velocity to the Tsallis distribution Eq. (1).
According to ref. [13], to include the radial flow in a
relativistic scenario, the Tsallis distribution function has been
expanded in a Taylor series in view of ($q-1$) being very small.
The normalized functional form of the distribution up to first
order in ($q-1$) is given by
\begin{align}
f_S(p_T)&=C_0 \bigg\{ 2T_0[rI_0(s)K_1(r)-sI_1(s)K_0(r)] \nonumber\\
&-(q-1)T_0r^2I_0(s)[K_0(r)+K_2(r)] \nonumber\\
&+4(q-1)T_0rsI_1(s)K_1(r) \nonumber\\
&-(q-1)T_0s^2K_0(r)[I_0(s)+I_2(s)] \nonumber\\
&+\frac{(q-1)}{4}T_0r^3I_0(s)[K_3(r)+3K_1(r)] \nonumber\\
&-\frac{3(q-1)}{2}T_0r^2s[K_2(r)+K_0(r)]I_1(s) \nonumber\\
&+\frac{3(q-1)}{2}T_0s^2r[I_0(s)+I_2(s)]K_1(r) \nonumber\\
&-\frac{(q-1)}{4}T_0s^3[I_3(s)+3I_1(s)]K_0(r) \bigg\},
\end{align}
where $C_0$ is the normalized constant which results in
$\int_0^{\infty} f_S(p_T)dp_T=1$, $r\equiv \gamma m_T/T_0$,
$s\equiv \gamma \beta_T p_T/T_0$, $\gamma=1/\sqrt{1-\beta_T^2}$,
and $I_n(s)$ and $K_n(r)$ are the modified Bessel functions of the
first and second kinds, respectively. We call Eq. (6) the improved
Tsallis distribution, which is in fact the Tsallis distribution
with radial flow, in which there are three free parameters
involved namely $T_0$, $q$, and $\beta_T$.

In most cases, the $p_T$ spectra are given in a wide $p_T$ range.
The improved Tsallis distribution, Eq. (6), is not enough to give
a good description. That is, the contribution of the hard
scattering process which occurs between valence quarks has to be
considered. We can use the inverse power-law
\begin{equation}
f_H(p_T)=Ap_T \bigg( 1+\frac{p_T}{ p_0} \bigg)^{-n},
\end{equation}
to describe the contribution of the hard scattering process, where
$p_0$ and $n$ are free parameters, and $A$ is the normalized
constant which depends on $p_0$ and $n$ and results in
$\int_0^{\infty} f_H(p_T)dp_T=1$. Eq. (7) results from the QCD
(quantum chromodynamics) calculus [30--32]. To describe the $p_T$
spectra in a wide $p_T$ range, we can use a superposition of the
improved Tsallis distribution which describes the contribution of
the soft excitation process and the inverse power-law which
describes the contribution of the hard scattering process
\begin{equation}
f_0(p_T)=kf_S(p_T)+(1-k)f_H(p_T),
\end{equation}
where $k$ denotes the contribution ratio (relative contribution or
fraction) of the improved Tsallis distribution and results
naturally in $\int_0^{\infty} f_0(p_T)dp_T=1$.

It should be noted that the above formalism describes the soft
component using the Tsallis distribution taking into account flow
in an approximate manner. For the hard component, only the
power-law distribution is used. We do not need to modify the
distribution for the hard component when taking into account the
flow due to the hard component being contributed from hard
scattering in the early stages of the collision when the flow is
not yet appearing. Comparatively, the soft component is
contributed from the soft excitation in the middle and later
stages of the collision when the flow is already appearing.
Although both the soft and hard components are power-law
distributions and they have a similar mathematical form before
taking into account the flow for the soft component, the
definitive final descriptions of the two components have different
forms.
\\

{\section{Results and discussion}}

Figure 1 presents the transverse momentum spectra, $(1/N_{EV})
(2\pi p_T)^{-1} d^2N/(dydp_T)$, of (a)-(c) $\pi^+$, $K^+$, and
$p$, as well as (b) $\pi^-$, $K^-$, and $\bar p$ produced in
(a)-(b) 0--5\% and (c) 70--80\% Au-Au collisions at
$\sqrt{s_{NN}}=14.5$ GeV, where $N_{EV}$ on the vertical axis
denotes the number of events, which means the mentioned quantity
[$(2\pi p_T)^{-1} d^2N/(dydp_T)$] per event and is usually omitted
in most cases, and $N$ denotes the number of particles. The
symbols represent the experimental data of the STAR Collaboration
measured in the rapidity range $|y|<0.1$ [21]. The solid curves
are our results calculated by using the improved Tsallis
distribution [13, 14]. The values of free parameters $T_0$, $q$,
and $\beta_T$, normalization constant $N_0$ which is used to fit
the data, and $\chi^2$ per degree of freedom ($\chi^2$/dof) are
listed in Table 1. One can see that the improved Tsallis
distribution describes the $p_T$ spectra of identified particles
produced in central (0--5\%) and peripheral (70--80\%) Au-Au
collisions at $\sqrt{s_{NN}}=14.5$ GeV.

Figure 2 is the same as Figure 1, but it shows the spectra, $(2\pi
p_T)^{-1} d^2N/(dydp_T)$, for (a)-(c) $\pi^+$, $K^+$, and $p$, as
well as (b)-(d) $\pi^-$, $K^-$, and $\bar p$ produced in (a)-(b)
0--5\% and (c)-(d) 70--80\% Au-Au collisions at
$\sqrt{s_{NN}}=62.4$ GeV. The experimental data of the STAR
Collaboration are taken from ref. [22]. One can see that the
improved Tsallis distribution describes the $p_T$ spectra of
identified particles produced in central (0--5\%) and peripheral
(70--80\%) Au-Au collisions at $\sqrt{s_{NN}}=62.4$ GeV.

In Figure 3, the transverse momentum spectra of (a)-(c) $\pi^+$,
$K^+$, and $p$, as well as (b)-(d) $\pi^-$, $K^-$, and $\bar p$
produced in (a)-(b) 0--5\% and (c)-(d) 60--92\% Au-Au collisions
at $\sqrt{s_{NN}}=130$ GeV are given. The symbols represent the
experimental data of the PHENIX Collaboration measured in the
pseudorapidity range $|\eta|<0.35$ [23, 24] and scaled by
different amounts shown in the panels. The solid, dotted, and
dashed curves are our results calculated by using the improved
Tsallis distribution [13, 14], the inverse power-law [30--32], and
their superposition, respectively. The values of free parameters
$T_0$, $q$, $\beta_T$, $k$, $p_0$, and $n$, normalization constant
$N_0$, and $\chi^2$/dof are listed in Table 2. One can see that in
most cases the superposition of the improved Tsallis distribution
and the inverse power-law describes the $p_T$ spectra of
identified particles produced in central (0--5\%) and peripheral
(60--92\%) Au-Au collisions at $\sqrt{s_{NN}}=130$ GeV.

The situation of Figure 4 is the same as Figure 3, but it shows
the spectra for (a)-(c) $\pi^+$, $K^+$, and $p$, as well as
(b)-(d) $\pi^-$, $K^-$, and $\bar p$ produced in (a)-(b) 0--5\%
and (c)-(d) 80--92\% Au-Au collisions at $\sqrt{s_{NN}}=200$ GeV.
The experimental data of the PHENIX Collaboration are taken from
refs. [18, 25--28]. One can see again that the superposition of
the improved Tsallis distribution and the inverse power-law
describes the $p_T$ spectra of identified particles produced in
central (0--5\%) and peripheral (80--92\%) Au-Au collisions at
$\sqrt{s_{NN}}=200$ GeV.

The situation of Figure 5 is also the same as Figure 3, but it
shows the spectra for (a)-(c) $\pi^+$, $K^+$, and $p$, as well as
(b)-(d) $\pi^-$, $K^-$, and $\bar p$ produced in (a)-(b) 5--10\%
and (c)-(d) 60--92\% Au-Au collisions at $\sqrt{s_{NN}}=200$ GeV.
The experimental data of the PHENIX Collaboration are taken from
refs. [18, 25--28]. Once more the superposition of the improved
Tsallis distribution and the inverse power-law describes the $p_T$
spectra of identified particles produced in central (5--10\%) and
peripheral (60--92\%) Au-Au collisions at $\sqrt{s_{NN}}=200$ GeV.

Figure 6 is the same as Figure 3, but it shows the spectra for
$\pi^++\pi^-$, $K^++K^-$, and $p+\bar p$ produced in (a) 0--5\%
and (b) 60--80\% Pb-Pb collisions at $\sqrt{s_{NN}}=2.76$ TeV. The
experimental data of the ALICE Collaboration are taken from ref.
[29] and measured in $|\eta|<0.8$ for high $p_T$ region and
$|y|<0.5$ for low $p_T$ region. Indeed, the superposition of the
improved Tsallis distribution and the inverse power-law describes
the $p_T$ spectra of identified particles produced in central
(0--5\%) and peripheral (60--80\%) Pb-Pb collisions at
$\sqrt{s_{NN}}=2.76$ TeV.

\begin{figure*}
\hskip-1.0cm
\begin{center}
\includegraphics[width=16.0cm]{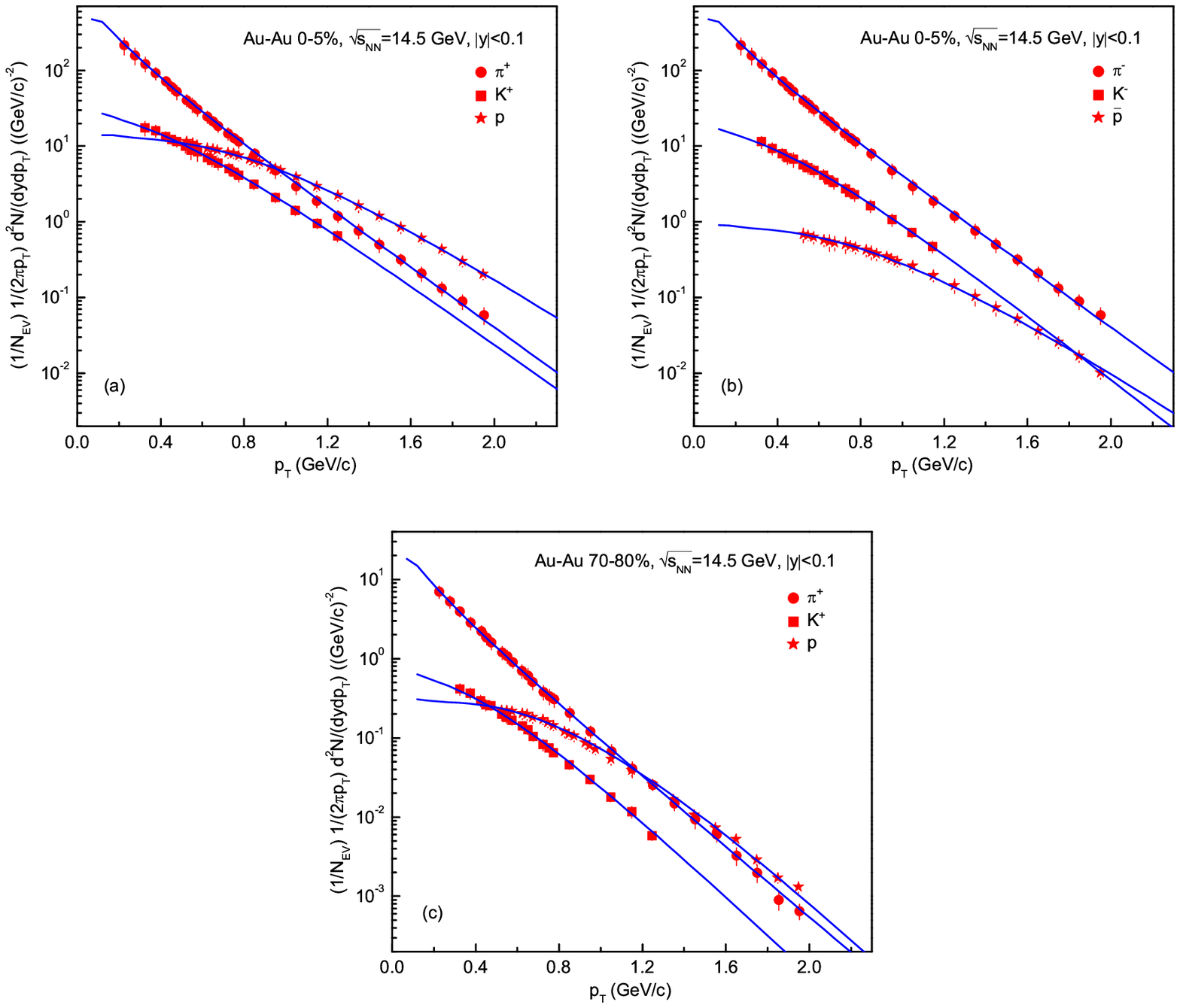}
\end{center}
\vskip0.5cm {\small Fig. 1. Transverse momentum spectra of (a)-(c)
$\pi^+$, $K^+$, and $p$, as well as (b) $\pi^-$, $K^-$, and $\bar
p$ produced in (a)-(b) 0--5\% and (c) 70--80\% Au-Au collisions at
$\sqrt{s_{NN}}=14.5$ GeV, where $N_{EV}$ on the vertical axis
denotes the number of events, which is usually omitted. The
symbols represent the experimental data of the STAR Collaboration
measured in the rapidity range $|y|<0.1$ [21]. The solid curves
are our results calculated by using the improved Tsallis
distribution [13, 14].}
\end{figure*}

\begin{figure*}
\hskip-1.0cm \begin{center}
\includegraphics[width=16.0cm]{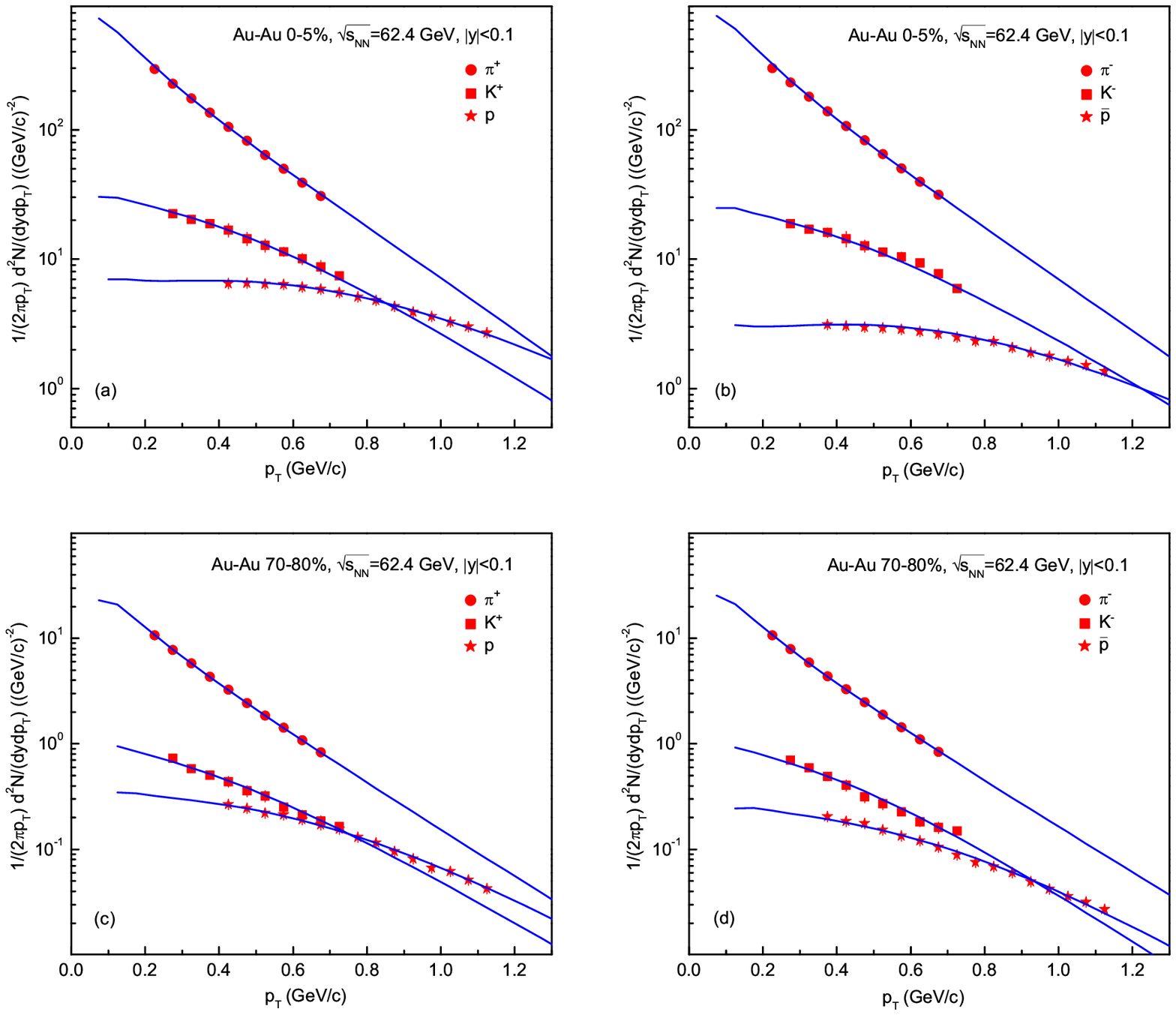}
\end{center}
\vskip0.5cm {\small Fig. 2. Same as Figure 1, but showing the
spectra for (a)-(c) $\pi^+$, $K^+$, and $p$, as well as (b)-(d)
$\pi^-$, $K^-$, and $\bar p$ produced in (a)-(b) 0--5\% and
(c)-(d) 70--80\% Au-Au collisions at $\sqrt{s_{NN}}=62.4$ GeV. The
experimental data of the STAR Collaboration are taken from ref.
[22].}
\end{figure*}

\begin{figure*}
\hskip-1.0cm \begin{center}
\includegraphics[width=16.0cm]{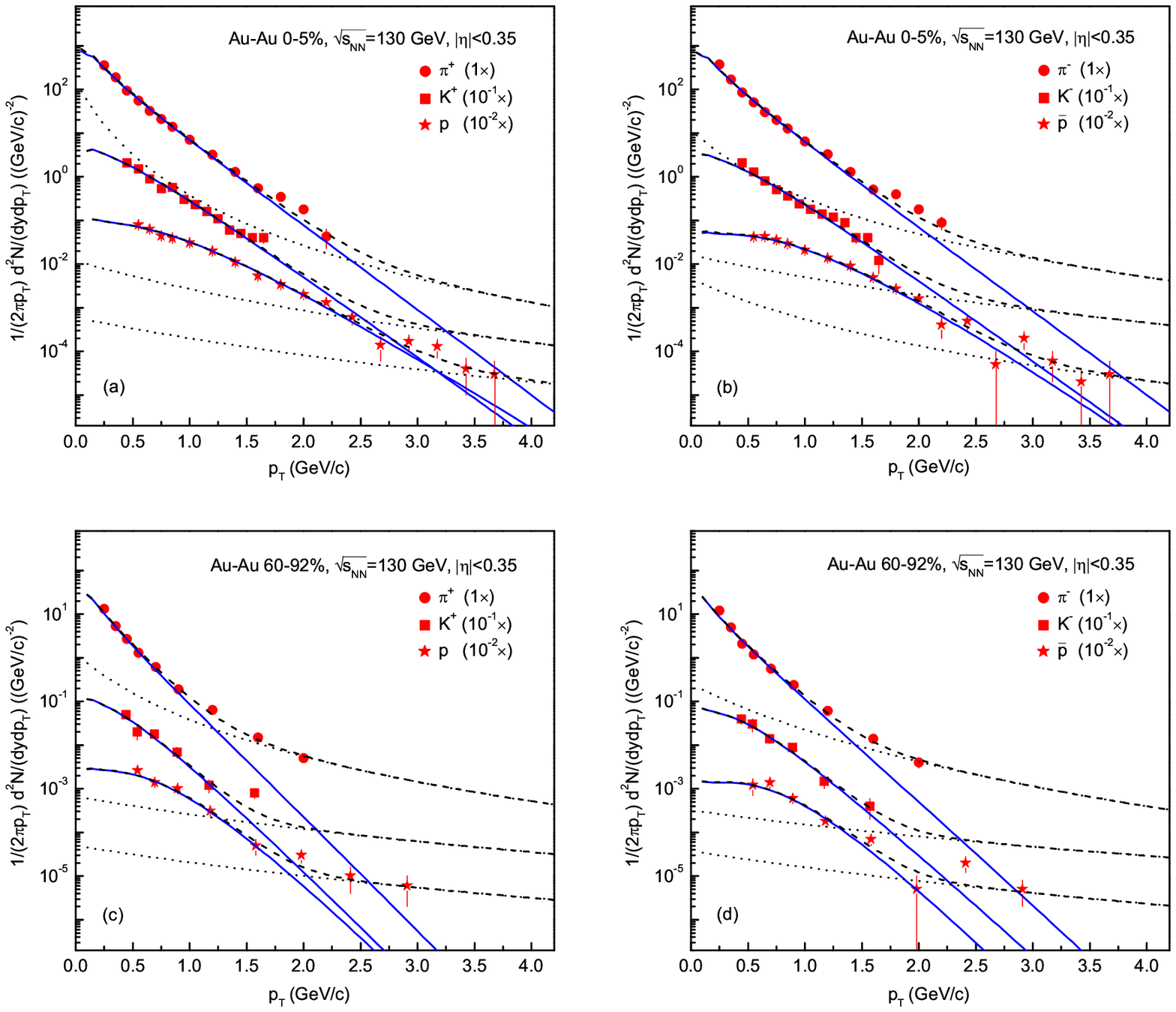}
\end{center}
\vskip0.5cm {\small Fig. 3. Transverse momentum spectra of (a)-(c)
$\pi^+$, $K^+$, and $p$, as well as (b)-(d) $\pi^-$, $K^-$, and
$\bar p$ produced in (a)-(b) 0--5\% and (c)-(d) 60--92\% Au-Au
collisions at $\sqrt{s_{NN}}=130$ GeV. The symbols represent the
experimental data of the PHENIX Collaboration measured in
$|\eta|<0.35$ [23, 24] and scaled by different amounts shown in
the panels. The solid, dotted, and dashed curves are our results
calculated by using the improved Tsallis distribution [13, 14],
the inverse power-law [30--32], and their superposition,
respectively.}
\end{figure*}

\begin{figure*}
\hskip-1.0cm \begin{center}
\includegraphics[width=16.0cm]{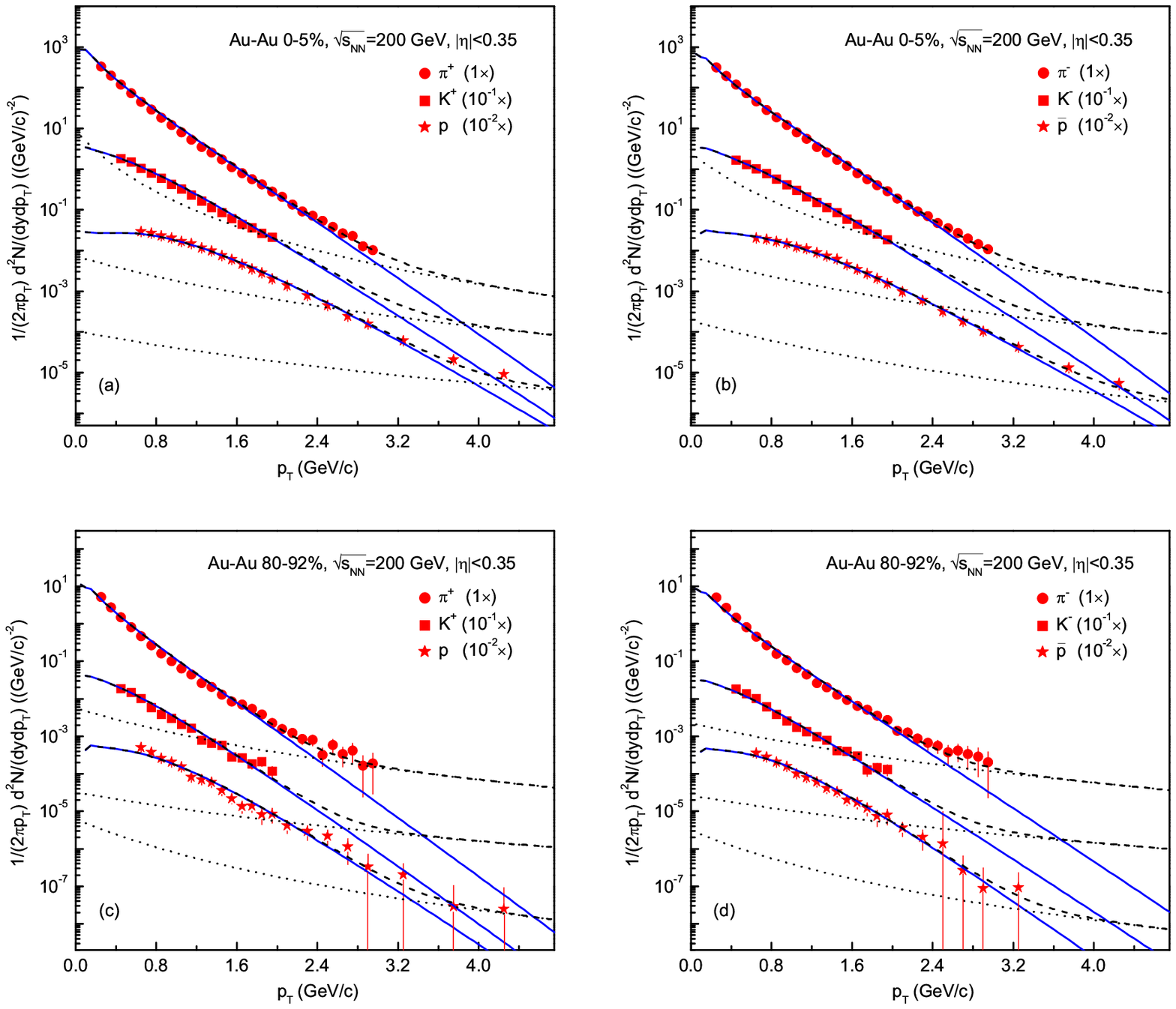}
\end{center}
\vskip0.5cm {\small Fig. 4. Same as Figure 3, but showing the
spectra for (a)-(c) $\pi^+$, $K^+$, and $p$, as well as (b)-(d)
$\pi^-$, $K^-$, and $\bar p$ produced in (a)-(b) 0--5\% and
(c)-(d) 80--92\% Au-Au collisions at $\sqrt{s_{NN}}=200$ GeV. The
experimental data of the PHENIX Collaboration are taken from refs.
[18, 25--28].}
\end{figure*}

\begin{figure*}
\hskip-1.0cm \begin{center}
\includegraphics[width=16.0cm]{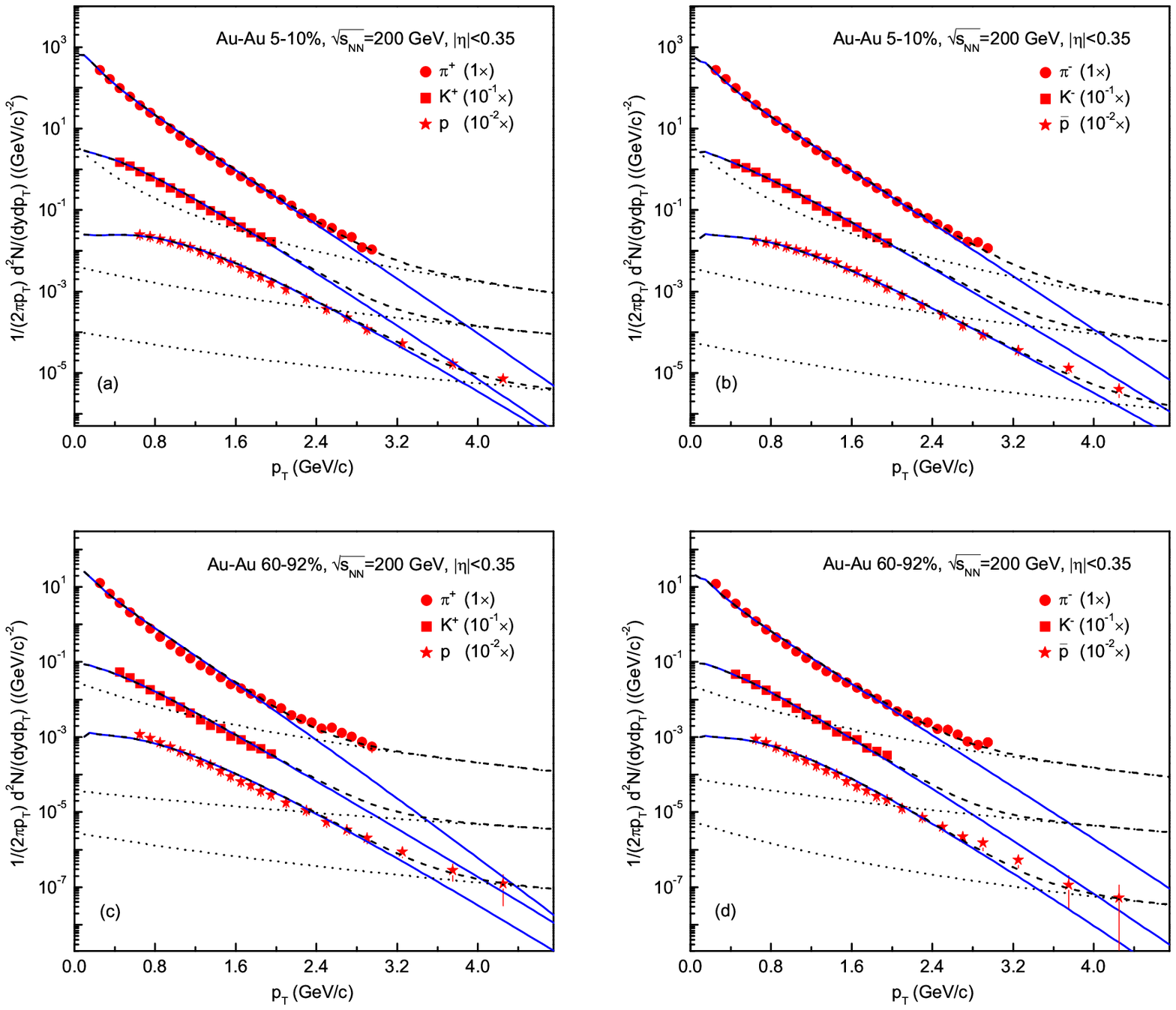}
\end{center}
\vskip0.5cm {\small Fig. 5. Same as Figure 3, but showing the
spectra for (a)-(c) $\pi^+$, $K^+$, and $p$, as well as (b)-(d)
$\pi^-$, $K^-$, and $\bar p$ produced in (a)-(b) 5--10\% and
(c)-(d) 60--92\% Au-Au collisions at $\sqrt{s_{NN}}=200$ GeV. The
experimental data of the PHENIX Collaboration are taken from refs.
[18, 25--28].}
\end{figure*}

\begin{figure*}
\hskip-1.0cm \begin{center}
\includegraphics[width=16.0cm]{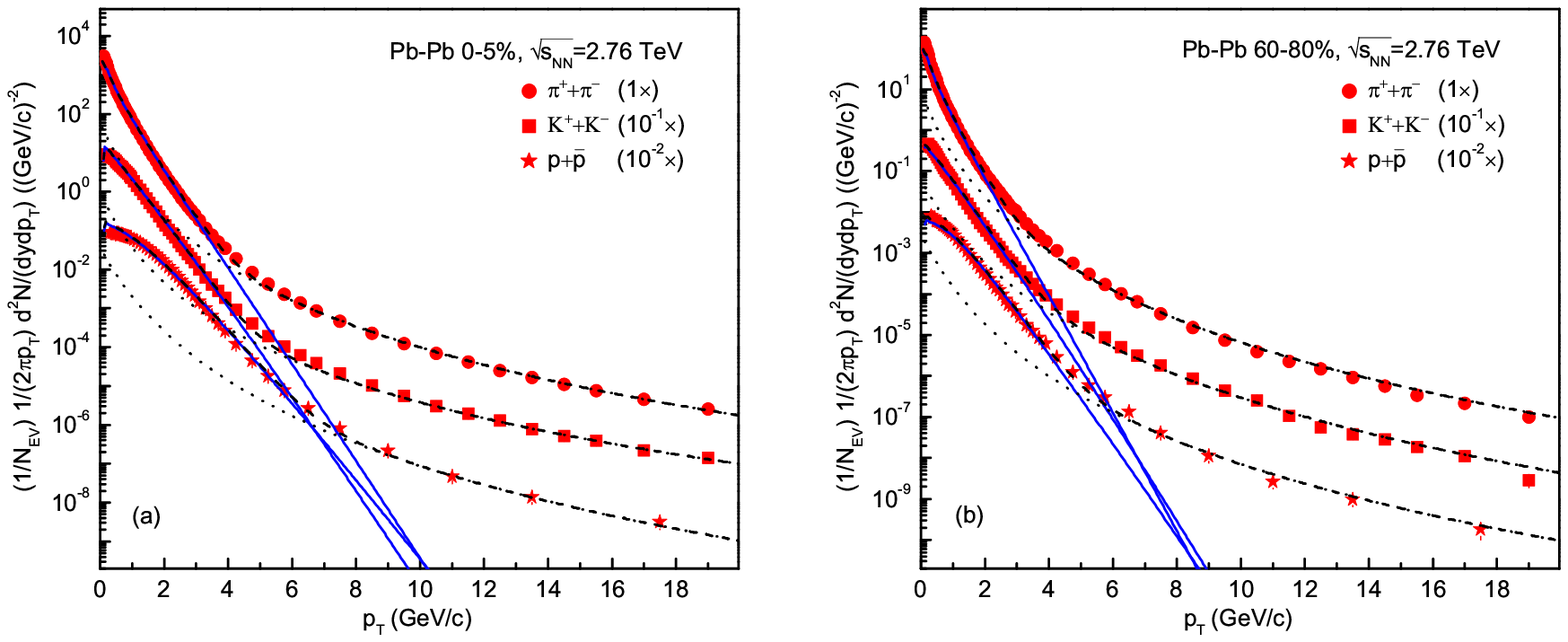}
\end{center}
\vskip0.5cm {\small Fig. 6. Same as Figure 3, but showing the
spectra for $\pi^++\pi^-$, $K^++K^-$, and $p+\bar p$ produced in
(a) 0--5\% and (b) 60--80\% Pb-Pb collisions at
$\sqrt{s_{NN}}=2.76$ TeV. The experimental data of the ALICE
Collaboration are taken from ref. [29] and measured in
$|\eta|<0.8$ for high $p_T$ region and $|y|<0.5$ for low $p_T$
region.}
\end{figure*}

\begin{table*}
{\small Table 1. Values of free parameters ($T_0$, $q$, and
$\beta_{T}$), normalization constant ($N_{0}$), and $\chi^2$/dof
corresponding to the curves in Figures 1 and 2.
\begin{center}
\begin{tabular}{cccccccc}
\hline\hline  Figure & Centrality & Particle & $T_0$ (GeV) & $q$ & $\beta_{T}$ ($c$) & $N_0$ & $\chi^2$/dof \\
\hline
1(a) &  0--5\%   & $\pi^+$   & $0.102\pm0.004$ & $1.011\pm0.005$  & $0.602\pm0.023$ & $108.589\pm13.114$ & $0.330$ \\
     &           & $K^+$     & $0.109\pm0.011$ & $1.002\pm0.001$  & $0.574\pm0.032$ & $10.606\pm1.683$   & $0.974$ \\
     &           & $p$       & $0.118\pm0.009$ & $1.003\pm0.002$  & $0.538\pm0.023$ & $10.525\pm1.213$   & $0.510$ \\
1(b) &  0--5\%   &$\pi^-$    & $0.102\pm0.004$ & $1.011\pm0.005$  & $0.602\pm0.023$ & $108.589\pm13.114$ & $0.330$ \\
     &           &$K^-$      & $0.102\pm0.009$ & $1.002\pm0.001$  & $0.558\pm0.032$ & $6.333\pm1.017$    & $0.469$ \\
     &           & $\bar{p}$ & $0.115\pm0.011$ & $1.005\pm0.003$  & $0.531\pm0.027$ & $0.665\pm0.053$    & $0.448$ \\
1(c) & 70--80\%  & $\pi^+$   & $0.091\pm0.004$ & $1.010\pm0.008$  & $0.604\pm0.030$ & $3.621\pm0.628$    & $0.248$ \\
     &           & $K^+$     & $0.091\pm0.009$ & $1.002\pm0.001$  & $0.535\pm0.030$ & $0.224\pm0.040$    & $1.071$ \\
     &           & $p$       & $0.091\pm0.007$ & $1.002\pm0.001$  & $0.499\pm0.018$ & $0.208\pm0.029$    & $1.057$ \\
\hline
2(a) &  0--5\%   & $\pi^+$   & $0.088\pm0.009$ & $1.089\pm0.055$  & $0.605\pm0.045$ & $152.903\pm11.752$ & $1.907$ \\
     &           & $K^+$     & $0.093\pm0.012$ & $1.049\pm0.039$  & $0.599\pm0.028$ & $14.564\pm1.169$   & $0.625$ \\
     &           & $p$       & $0.110\pm0.013$ & $1.017\pm0.015$  & $0.588\pm0.025$ & $6.838\pm0.388$    & $4.341$ \\
2(b) &  0--5\%   &$\pi^-$    & $0.091\pm0.013$ & $1.062\pm0.032$  & $0.604\pm0.038$ & $159.261\pm16.224$ & $3.201$ \\
     &           &$K^-$      & $0.095\pm0.012$ & $1.051\pm0.041$  & $0.604\pm0.028$ & $12.253\pm1.051$   & $2.051$ \\
     &           &$\bar{p}$  & $0.116\pm0.011$ & $1.002\pm0.001$  & $0.601\pm0.014$ & $3.113\pm0.183$    & $9.118$ \\
2(c) &  70--80\% & $\pi^+$   & $0.099\pm0.009$ & $1.005\pm0.004$  & $0.588\pm0.038$ & $5.061\pm0.622$    & $0.495$ \\
     &           & $K^+$     & $0.104\pm0.011$ & $1.001\pm0.0008$ & $0.551\pm0.028$ & $0.350\pm0.030$    & $1.866$ \\
     &           & $p$       & $0.104\pm0.011$ & $1.001\pm0.0008$ & $0.474\pm0.025$ & $0.209\pm0.022$    & $2.025$ \\
2(d) &  70--80\% &$\pi^-$    & $0.095\pm0.009$ & $1.015\pm0.013$  & $0.596\pm0.038$ & $5.233\pm0.579$    & $0.460$ \\
     &           &$K^-$      & $0.095\pm0.008$ & $1.001\pm0.0008$ & $0.534\pm0.029$ & $0.325\pm0.040$    & $2.651$ \\
     &           &$\bar{p}$  & $0.103\pm0.011$ & $1.001\pm0.0008$ & $0.449\pm0.021$ & $0.140\pm0.015$    & $4.580$ \\
\hline
\end{tabular}
\end{center}}
\end{table*}

\begin{table*}
{\small Table 2. Values of free parameters ($T_0$, $q$,
$\beta_{T}$, $k$, $p_{0}$, and $n$), normalization constant
($N_{0}$), and $\chi^2$/dof corresponding to the curves in Figures
3--6. \tiny
\begin{center}
\begin{tabular}{ccccccccccc}
\hline\hline Figure & Centrality & Particle & $T_0$ (GeV) & $q$ & $\beta_{T}$ ($c$) & $k$ & $p_{0}$ (GeV/$c$) & $n$ & $N_0$ & $\chi^2$/dof \\
\hline
3(a) & 0--5\%   & $\pi^+$       & $0.109\pm0.006$ & $1.019\pm0.015$ & $0.558\pm0.045$ & $0.945\pm0.052$ & $0.425\pm0.072$ & $4.994\pm0.912$ & $195.464\pm37.486$  & $12.415$\\
     &          & $K^+$         & $0.115\pm0.009$ & $1.017\pm0.015$ & $0.537\pm0.025$ & $0.988\pm0.041$ & $2.805\pm0.662$ & $4.871\pm0.072$ & $17.615\pm3.339$    & $2.620$\\
     &          & $p$           & $0.131\pm0.014$ & $1.021\pm0.020$ & $0.531\pm0.031$ & $0.982\pm0.015$ & $3.988\pm1.212$ & $4.872\pm0.895$ & $7.311\pm1.463$     & $1.435$\\
3(b) & 0--5\%   & $\pi^-$       & $0.111\pm0.005$ & $1.010\pm0.008$ & $0.568\pm0.025$ & $0.976\pm0.028$ & $0.705\pm0.082$ & $4.105\pm0.512$ & $170.329\pm26.755$  & $18.306$\\
     &          & $K^-$         & $0.111\pm0.009$ & $1.012\pm0.010$ & $0.568\pm0.029$ & $0.965\pm0.031$ & $3.822\pm0.454$ & $4.964\pm0.712$ & $13.713\pm1.436$    & $2.435$\\
     &          & $\bar{p}$     & $0.111\pm0.012$ & $1.066\pm0.052$ & $0.568\pm0.034$ & $0.951\pm0.041$ & $1.407\pm0.291$ & $4.015\pm0.812$ & $4.786\pm0.801$     & $1.494$\\
3(c) & 60--92\% & $\pi^+$       & $0.090\pm0.009$ & $1.022\pm0.021$ & $0.481\pm0.045$ & $0.917\pm0.015$ & $0.821\pm0.135$ & $4.408\pm0.442$ & $5.165\pm0.603$     & $18.640$\\
     &          & $K^+$         & $0.090\pm0.013$ & $1.015\pm0.014$ & $0.481\pm0.051$ & $0.901\pm0.045$ & $3.821\pm0.632$ & $4.104\pm0.972$ & $0.396\pm0.053$     & $8.166^*$\\
     &          & $p$           & $0.090\pm0.015$ & $1.005\pm0.004$ & $0.481\pm0.049$ & $0.884\pm0.021$ & $4.449\pm0.542$ & $4.280\pm0.545$ & $0.196\pm0.030$     & $3.852$\\
3(d) & 60--92\% & $\pi^-$       & $0.085\pm0.009$ & $1.021\pm0.015$ & $0.588\pm0.035$ & $0.952\pm0.024$ & $1.909\pm0.045$ & $5.697\pm0.595$ & $4.413\pm0.539$     & $40.442$\\
     &          & $K^-$         & $0.085\pm0.009$ & $1.014\pm0.013$ & $0.588\pm0.044$ & $0.908\pm0.085$ & $4.955\pm0.958$ & $4.068\pm0.988$ & $0.294\pm0.060$     & $2.931^*$\\
     &          & $\bar{p}$     & $0.085\pm0.009$ & $1.007\pm0.006$ & $0.512\pm0.049$ & $0.869\pm0.071$ & $5.368\pm0.682$ & $4.986\pm0.512$ & $0.115\pm0.019$     & $9.566$\\
\hline
4(a) & 0--5\%   & $\pi^+$       & $0.109\pm0.006$ & $1.023\pm0.016$ & $0.634\pm0.019$ & $0.991\pm0.008$ & $0.709\pm0.165$ & $4.597\pm0.645$ & $224.148\pm44.212$  & $2.205$\\
     &          & $K^+$         & $0.113\pm0.011$ & $1.052\pm0.037$ & $0.634\pm0.029$ & $0.992\pm0.007$ & $2.594\pm0.323$ & $4.291\pm0.935$ & $16.743\pm2.746$    & $0.916$\\
     &          & $p$           & $0.121\pm0.009$ & $1.097\pm0.055$ & $0.634\pm0.021$ & $0.991\pm0.007$ & $3.697\pm0.442$ & $4.036\pm0.834$ & $3.191\pm0.477$     & $0.940$\\
4(b) & 0--5\%   & $\pi^-$       & $0.111\pm0.006$ & $1.062\pm0.025$ & $0.590\pm0.032$ & $0.994\pm0.005$ & $0.956\pm0.218$ & $4.436\pm0.585$ & $180.622\pm38.234$  & $1.529$\\
     &          & $K^-$         & $0.121\pm0.011$ & $1.067\pm0.049$ & $0.590\pm0.031$ & $0.991\pm0.004$ & $2.667\pm0.331$ & $4.286\pm0.635$ & $15.977\pm4.268$    & $1.087$\\
     &          & $\bar{p}$     & $0.133\pm0.009$ & $1.095\pm0.065$ & $0.590\pm0.028$ & $0.992\pm0.007$ & $3.132\pm0.921$ & $4.969\pm0.812$ & $2.777\pm0.564$     & $0.971$\\
4(c) & 80--92\% & $\pi^+$       & $0.106\pm0.009$ & $1.028\pm0.017$ & $0.572\pm0.026$ & $0.994\pm0.005$ & $2.321\pm0.461$ & $4.382\pm1.454$ & $2.731\pm0.449$     & $3.616$\\
     &          & $K^+$         & $0.106\pm0.008$ & $1.028\pm0.021$ & $0.572\pm0.055$ & $0.991\pm0.008$ & $3.802\pm0.395$ & $4.161\pm1.115$ & $0.175\pm0.032$     & $1.847$\\
     &          & $p$           & $0.119\pm0.009$ & $1.013\pm0.010$ & $0.528\pm0.035$ & $0.992\pm0.005$ & $1.914\pm0.258$ & $4.945\pm0.612$ & $0.041\pm0.003$     & $1.551$\\
4(d) & 80--92\% & $\pi^-$       & $0.115\pm0.007$ & $1.015\pm0.012$ & $0.587\pm0.039$ & $0.995\pm0.004$ & $3.419\pm0.423$ & $4.706\pm0.612$ & $2.161\pm0.323$     & $0.319$\\
     &          & $K^-$         & $0.115\pm0.009$ & $1.013\pm0.011$ & $0.587\pm0.039$ & $0.990\pm0.006$ & $3.922\pm0.458$ & $4.135\pm0.995$ & $0.137\pm0.022$     & $3.075$\\
     &          & $\bar{p}$     & $0.116\pm0.009$ & $1.005\pm0.004$ & $0.529\pm0.038$ & $0.995\pm0.004$ & $1.859\pm0.423$ & $4.796\pm0.825$ & $0.035\pm0.005$     & $0.523$\\
\hline
5(a) & 5--10\%  & $\pi^+$       & $0.111\pm0.005$ & $1.025\pm0.015$ & $0.635\pm0.023$ & $0.992\pm0.006$ & $0.742\pm0.085$ & $4.206\pm0.681$ & $168.336\pm20.667$  & $1.714$\\
     &          & $K^+$         & $0.111\pm0.009$ & $1.031\pm0.025$ & $0.635\pm0.026$ & $0.991\pm0.008$ & $3.191\pm0.295$ & $4.219\pm0.045$ & $13.864\pm2.680$    & $0.795$\\
     &          & $p$           & $0.118\pm0.009$ & $1.089\pm0.055$ & $0.635\pm0.019$ & $0.990\pm0.005$ & $4.271\pm0.318$ & $4.465\pm0.052$ & $2.872\pm0.587$     & $0.411$\\
5(b) & 5--10\%  & $\pi^-$       & $0.121\pm0.007$ & $1.026\pm0.018$ & $0.599\pm0.029$ & $0.992\pm0.006$ & $0.912\pm0.182$ & $4.973\pm1.112$ & $141.910\pm31.630$  & $1.155$\\
     &          & $K^-$         & $0.128\pm0.013$ & $1.039\pm0.026$ & $0.599\pm0.039$ & $0.993\pm0.006$ & $3.146\pm0.235$ & $4.546\pm1.485$ & $12.905\pm1.965$    & $1.051$\\
     &          & $\bar{p}$     & $0.139\pm0.009$ & $1.019\pm0.015$ & $0.599\pm0.019$ & $0.995\pm0.004$ & $3.695\pm0.544$ & $4.605\pm0.385$ & $2.336\pm0.452$     & $1.114$\\
5(c) & 60--92\% & $\pi^+$       & $0.098\pm0.007$ & $1.068\pm0.025$ & $0.608\pm0.019$ & $0.991\pm0.005$ & $1.912\pm0.852$ & $4.425\pm0.745$ & $5.295\pm1.115$     & $3.382$\\
     &          & $K^+$         & $0.121\pm0.011$ & $1.015\pm0.012$ & $0.608\pm0.036$ & $0.989\pm0.010$ & $5.689\pm0.995$ & $3.879\pm0.835$ & $0.395\pm0.207$     & $2.223$\\
     &          & $p$           & $0.126\pm0.009$ & $1.032\pm0.026$ & $0.558\pm0.019$ & $0.992\pm0.005$ & $4.578\pm0.985$ & $4.887\pm1.152$ & $0.103\pm0.011$     & $1.990$\\
5(d) & 60--92\% & $\pi^-$       & $0.114\pm0.006$ & $1.026\pm0.012$ & $0.599\pm0.035$ & $0.992\pm0.007$ & $2.188\pm0.242$ & $4.857\pm1.445$ & $5.244\pm0.970$     & $2.671$\\
     &          & $K^-$         & $0.114\pm0.006$ & $1.030\pm0.018$ & $0.575\pm0.038$ & $0.990\pm0.003$ & $4.393\pm0.812$ & $4.506\pm0.758$ & $0.403\pm0.055$     & $2.913$\\
     &          & $\bar{p}$     & $0.114\pm0.007$ & $1.026\pm0.021$ & $0.556\pm0.025$ & $0.994\pm0.005$ & $2.187\pm0.545$ & $4.475\pm0.915$ & $0.088\pm0.010$     & $1.128$\\
\hline
6(a) & 0--5\%   & $\pi^{\pm}$   & $0.161\pm0.009$ & $1.009\pm0.008$ & $0.612\pm0.034$ & $0.976\pm0.007$ & $1.440\pm0.144$ & $6.438\pm0.196$ & $667.706\pm118.159$ & $3.901$\\
     &          & $K^{\pm}$     & $0.162\pm0.009$ & $1.049\pm0.041$ & $0.612\pm0.033$ & $0.976\pm0.007$ & $1.272\pm0.129$ & $5.742\pm0.205$ & $72.995\pm11.077$   & $2.196$\\
     &          & $p+\bar{p}$   & $0.192\pm0.012$ & $1.075\pm0.055$ & $0.612\pm0.023$ & $0.961\pm0.011$ & $2.073\pm0.169$ & $7.349\pm0.251$ & $14.629\pm2.695$    & $6.755$\\
6(b) & 60--80\% & $\pi^{\pm}$   & $0.135\pm0.011$ & $1.011\pm0.010$ & $0.630\pm0.033$ & $0.914\pm0.021$ & $1.426\pm0.094$ & $6.779\pm0.146$ & $25.246\pm4.326$    & $6.775$\\
     &          & $K^{\pm}$     & $0.158\pm0.014$ & $1.048\pm0.041$ & $0.608\pm0.035$ & $0.921\pm0.019$ & $2.144\pm0.172$ & $7.045\pm0.198$ & $2.124\pm0.441$     & $4.190$\\
     &          & $p+\bar{p}$   & $0.176\pm0.015$ & $1.017\pm0.015$ & $0.598\pm0.039$ & $0.922\pm0.018$ & $1.899\pm0.045$ & $7.059\pm0.147$ & $0.611\pm0.105$     & $2.926$\\
\hline
\end{tabular}
\end{center}
$^*$ This is $\chi^2$ only due to the dof being less than 1. }
\end{table*}

\begin{figure*}
\hskip-1.0cm \begin{center}
\includegraphics[width=16.0cm]{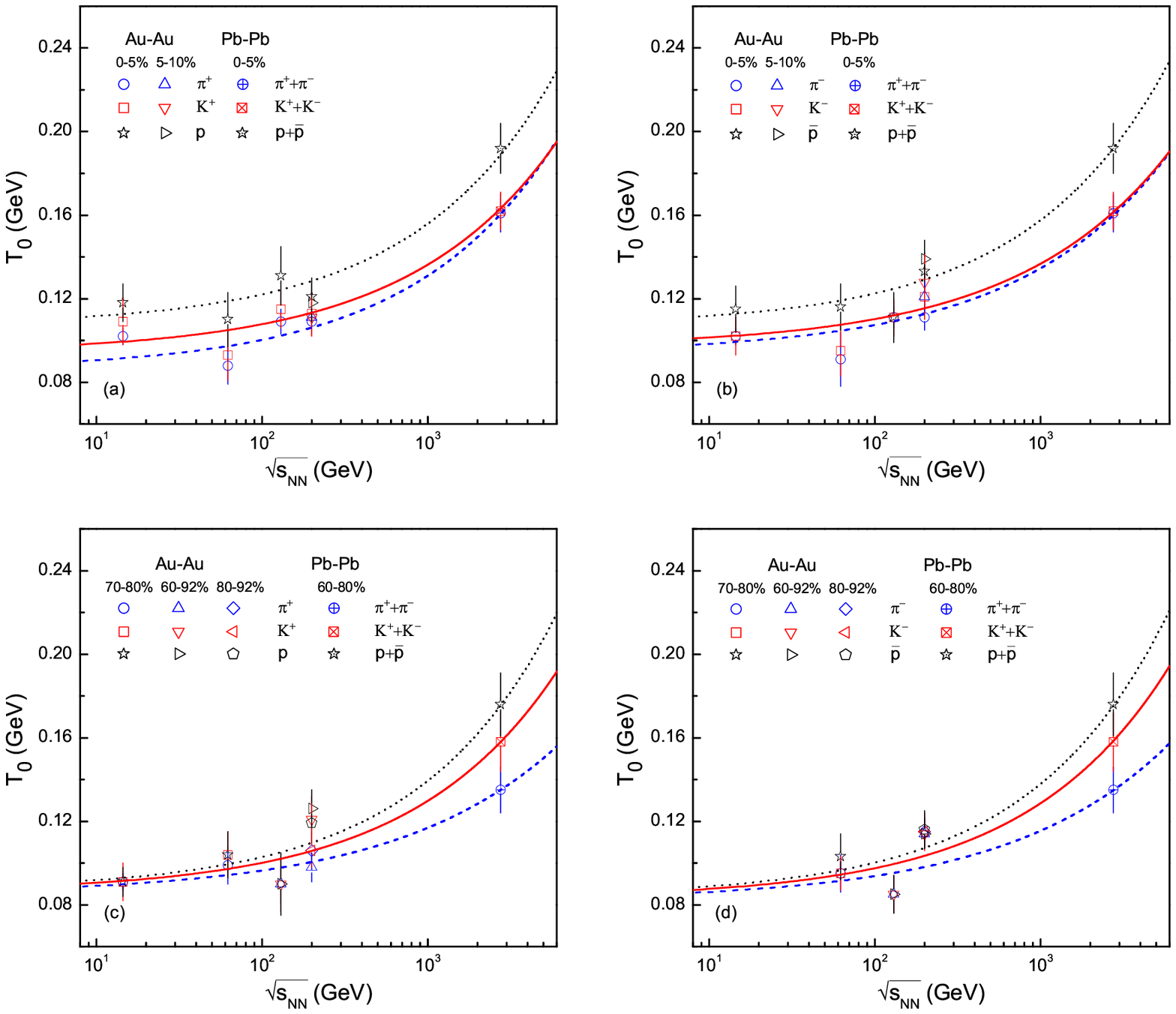}
\end{center}
\vskip0.5cm {\small Fig. 7. Dependences of $T_0$ on
$\sqrt{s_{NN}}$ for (a)-(c) positively and (b)-(d) negatively
charged particles in (a)-(b) central and (c)-(d) peripheral Au-Au
collisions at different energies, as well as for charged particles
in (a)-(b) central and (c)-(d) peripheral Pb-Pb collisions at 2.76
TeV. The symbols represent $T_0$ extracted from different spectra
for different particles in different collisions listed in Tables 1
and 2 and shown in the panels. The dashed, solid, and dotted
curves are our results fitted by using the method of least squares
for charged pions, kaons, and protons (antiprotons),
respectively.}
\end{figure*}

\begin{table*}
{\small Table 3. Values of parameters ($a$, $b$, and $c$) and
$\chi^2$/dof corresponding to the curves in Figure 7. The function
for the curves in Figure 7 is $T_0=a(\sqrt{s_{NN}})^b+c$.
\begin{center}
\begin{tabular}{cccccc}
\hline\hline  Figure & Main particle & $a$ & $b$ & $c$ & $\chi^2$/dof \\
\hline
7(a) & $\pi^+$  & $(1.442\pm0.115)\times10^{-3}$ & $0.498\pm0.021$ & $0.086\pm0.004$ & $3.479$ \\
     & $K^+$    & $(1.462\pm0.151)\times10^{-3}$ & $0.487\pm0.022$ & $0.094\pm0.003$ & $0.740$ \\
     & $p$      & $(1.432\pm0.235)\times10^{-3}$ & $0.511\pm0.019$ & $0.107\pm0.005$ & $1.078$ \\
7(b) & $\pi^-$  & $(1.462\pm0.252)\times10^{-3}$ & $0.481\pm0.029$ & $0.094\pm0.009$ & $1.048$ \\
     & $K^-$    & $(1.462\pm0.232)\times10^{-3}$ & $0.478\pm0.025$ & $0.097\pm0.007$ & $0.767$ \\
     & $\bar p$ & $(1.462\pm0.199)\times10^{-3}$ & $0.513\pm0.020$ & $0.107\pm0.005$ & $0.941$ \\
7(c) & $\pi^+$  & $(1.462\pm0.225)\times10^{-3}$ & $0.446\pm0.021$ & $0.085\pm0.005$ & $0.551$ \\
     & $K^+$    & $(1.462\pm0.235)\times10^{-3}$ & $0.492\pm0.015$ & $0.086\pm0.005$ & $1.051$ \\
     & $p$      & $(1.462\pm0.193)\times10^{-3}$ & $0.518\pm0.015$ & $0.087\pm0.006$ & $1.822$ \\
7(d) & $\pi^-$  & $(1.462\pm0.213)\times10^{-3}$ & $0.453\pm0.022$ & $0.082\pm0.004$ & $7.138$ \\
     & $K^-$    & $(1.462\pm0.202)\times10^{-3}$ & $0.498\pm0.013$ & $0.083\pm0.004$ & $3.669$ \\
     & $\bar p$ & $(1.462\pm0.165)\times10^{-3}$ & $0.522\pm0.015$ & $0.084\pm0.004$ & $2.973$ \\
\hline
\end{tabular}
\end{center}}
\end{table*}

To study the changing tendencies of parameters, Figure 7 shows the
dependences of $T_0$ on $\sqrt{s_{NN}}$ for (a)-(c) positively and
(b)-(d) negatively charged particles in (a)-(b) central and
(c)-(d) peripheral Au-Au collisions at different energies, as well
as for charged particles in (a)-(b) central and (c)-(d) peripheral
Pb-Pb collisions at 2.76 TeV. The symbols represent $T_0$
extracted from different spectra for different particles in
different collisions listed in Tables 1 and 2 and shown in the
panels. The dashed, solid, and dotted curves are our results
fitted by using the method of least squares for charged pions,
kaons, and protons (antiprotons), respectively. These curves are
described by the function
\begin{equation}
T_0=a(\sqrt{s_{NN}})^b+c,
\end{equation}
where the values of parameters $a$, $b$, and $c$, as well as
$\chi^2$/dof are given in Table 3. One can see that $T_0$
increases with the increase of $\sqrt{s_{NN}}$, the function
describes the tendency of $T_0$ in most cases.

The dependence of $T_0$ on $\sqrt{s_{NN}}$ obtained in the present
work is inconsistent with the original blast-wave model [11, 12,
20, 22] which gives a lower $T_0$ at higher energy. Although the
lower $T_0$ can be explained to a longer lifetime of the hot and
dense QGP, the higher $T_0$ can be explained to a higher
excitation degree. The present work extracted a higher $T_0$ in
central collisions than in peripheral collisions, which is
consistent with the improved blast-wave model [33] which uses
sources of particle emission from a Tsallis distribution, and
inconsistent with the original blast-wave model [11, 12, 20, 22]
which uses sources of particle emission from a Boltzmann
distribution. This difference can be also explained by the higher
excitation degree or longer lifetime, or different `thermometers'
or `thermometric scales' being used. From Figure 7, one can also
see the slight differences for different particles in some cases.
This confirms the mass-dependent differential kinetic freeze-out
scenario [14, 16].

Figures 8--14 are the same as Figure 7, but they show the
dependences of $q$, $\beta_T$, $p_0$, $n$, $k$, $kN_0$, and $N_0$
on $\sqrt{s_{NN}}$, respectively, where the product $kN_0$ in
Figure 13 represents the yield of soft excitation process. The
horizontal dashed, solid, and dotted lines in Figure 8 (or Figure
12) represent the mean values of $q$ (or $k$) over different
energies for charged pions, kaons, and protons (antiprotons),
respectively. The dashed, solid, and dotted curves are our results
fitted by using the method of least squares for charged pions,
kaons, and protons (antiprotons), respectively, though some curves
do not describe the tendencies of parameters. The function for the
curves in Figures 8--12 is
\begin{equation}
Y=a+b\ln(\sqrt{s_{NN}}),
\end{equation}
where $Y=q$, $\beta_T$, $p_0$, $n$, or $k$. The function for the
curves in Figures 13 and 14 is
\begin{equation}
Y=\exp[a+b\ln(\sqrt{s_{NN}})],
\end{equation}
where $Y=kN_0$ or $N_0$. The values of parameters $a$ and $b$, as
well as $\chi^2$/dof are given in Table 4. One can see that with
the increase of $\sqrt{s_{NN}}$, $q$ and $\beta_T$ increase
slightly, $p_0$ and $k$ decrease slightly, $n$, $kN_0$, and $N_0$
increase generally. The functions describe the tendencies of
parameters in some cases, while in other cases the functions fail
to describe the tendencies.

The parameter $q$ increases slightly with the increase of
$\sqrt{s_{NN}}$, but the dependence of $q$ on $\sqrt{s_{NN}}$ is
not obvious. As the entropy index, $q$ describes the degree
departing from the equilibrium state or the degree of
non-equilibrium. The parameters $q$ in central and peripheral
collisions are very small, which means that the two types of
collisions are in the nearly equilibrium state respectively,
though a slightly larger $q$ seems to be observed in central
collisions. In most cases, the differences in $q$ for different
particles are not obvious.

The parameter $\beta_T$ increases slightly with the increase of
$\sqrt{s_{NN}}$. Moreover, the present work also extracted a
slightly larger $\beta_T$ in central collisions than in peripheral
collisions, which is partly inconsistent with the blast-wave model
which gives non-zero flow velocity in central collisions and zero
flow velocity in peripheral collisions [11, 12, 20, 22, 33]. The
present work confirms our recent work [17] which used the
alternative method and obtained a slightly larger $\beta_T$ in
central collisions than in peripheral collisions, though the
values obtained in the present work are slightly larger than those
in our recent work. In our opinion, the flow is produced in the
inner core of the interacting system. Even for peripheral or
proton-proton collisions, there is non-zero flow velocity. From
Figure 9, one can also see the mass-dependence of $\beta_T$ in
most cases. A heavy particle corresponds to a small $\beta_T$ due
to its large inertia. The differences in $\beta_T$ for different
particles decrease with the increase of $\sqrt{s_{NN}}$, where
$\beta_T$ is large at high $\sqrt{s_{NN}}$. This reflects the fact
that the mass effect can be neglected in a strong flow field.

\begin{figure*}
\hskip-1.0cm \begin{center}
\includegraphics[width=16.0cm]{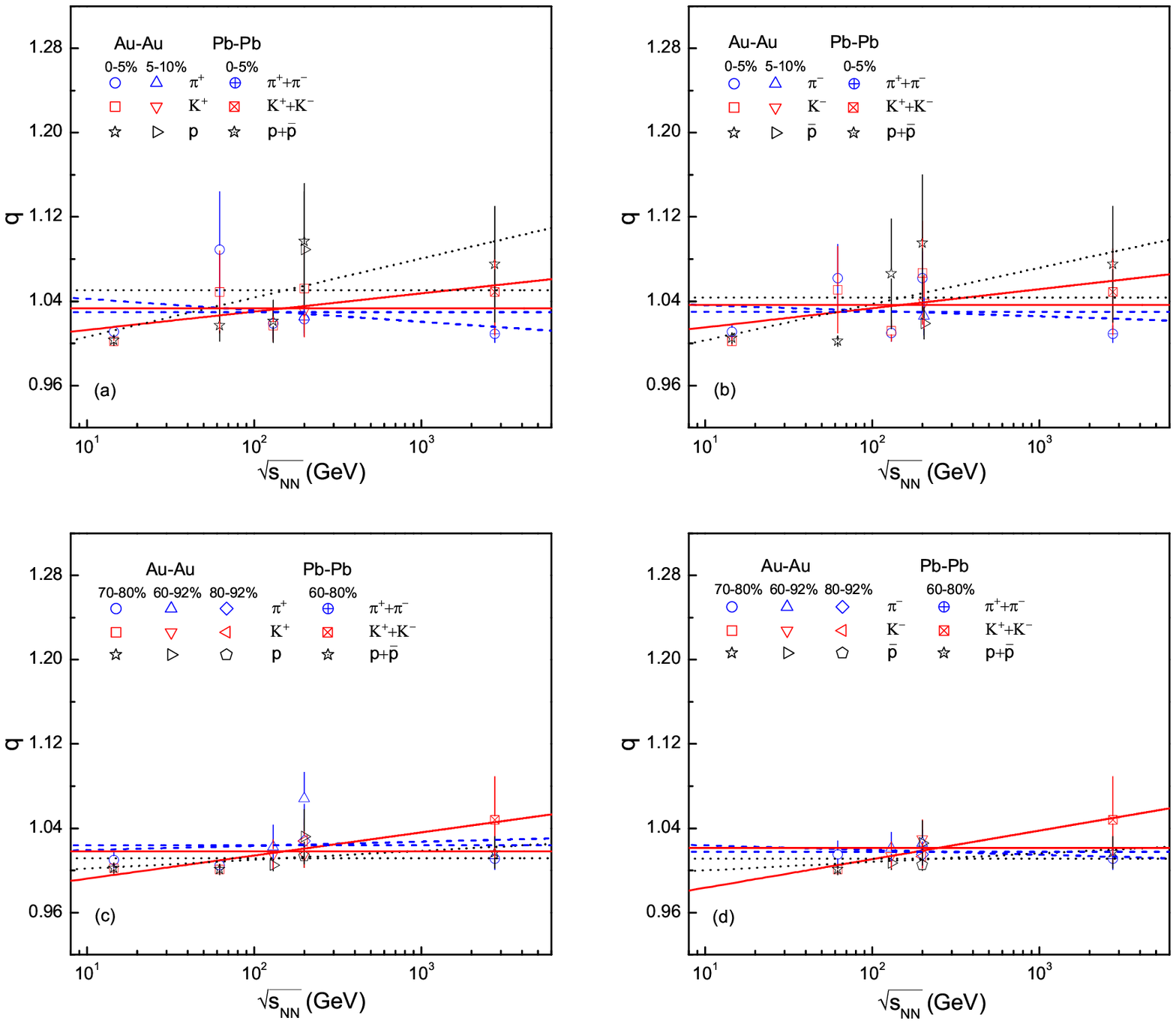}
\end{center}
\vskip0.5cm {\small Fig. 8. Same as Figure 7, but showing the
dependences of $q$ on $\sqrt{s_{NN}}$. The horizontal dashed,
solid, and dotted lines represent the mean values of $q$ over
different energies for charged pions, kaons, and protons
(antiprotons), respectively.}
\end{figure*}

\begin{figure*}
\hskip-1.0cm \begin{center}
\includegraphics[width=16.0cm]{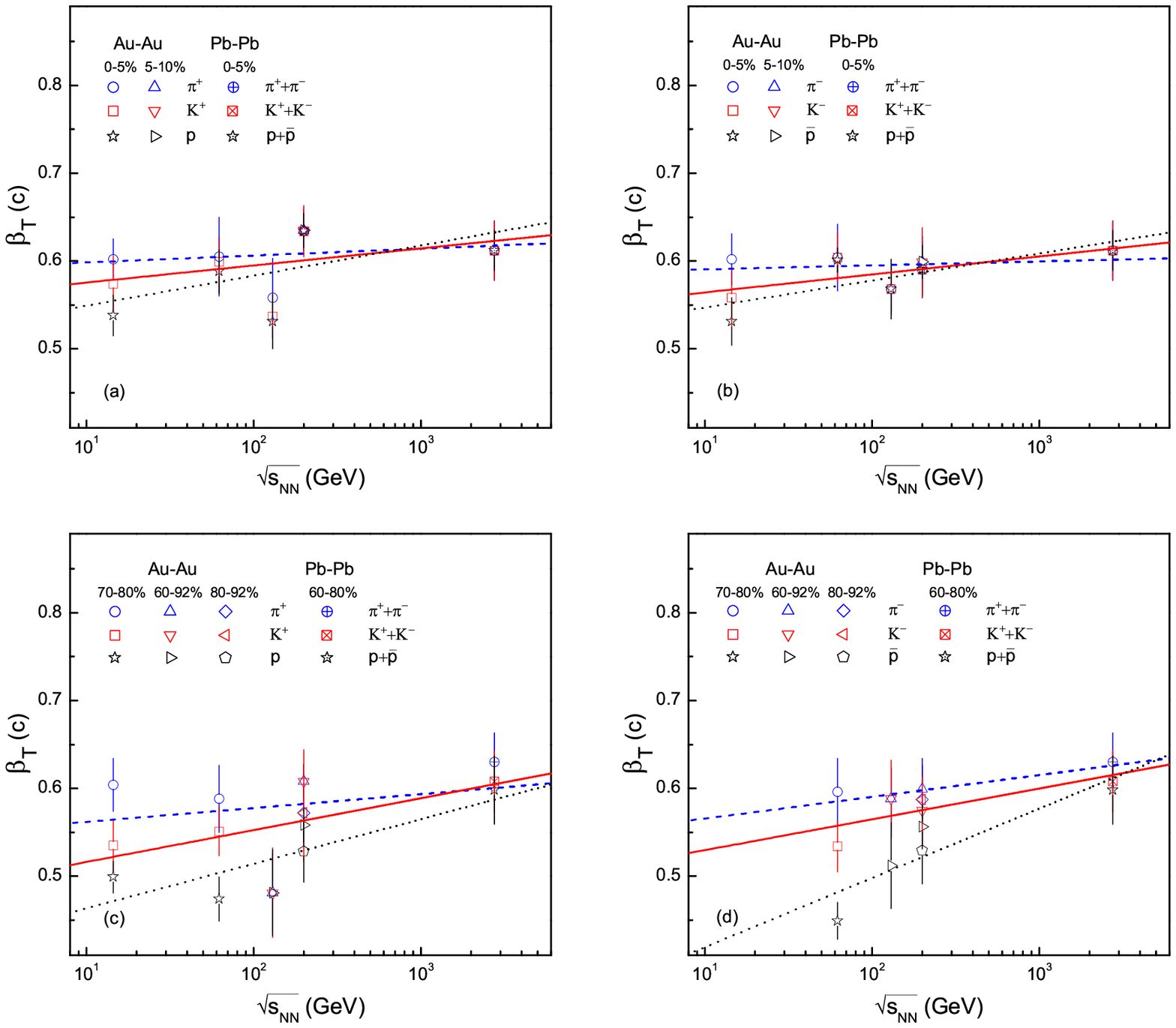}
\end{center}
\vskip0.5cm {\small Fig. 9. Same as Figure 7, but showing the
dependences of $\beta_T$ on $\sqrt{s_{NN}}$.}
\end{figure*}

\begin{figure*}
\hskip-1.0cm \begin{center}
\includegraphics[width=16.0cm]{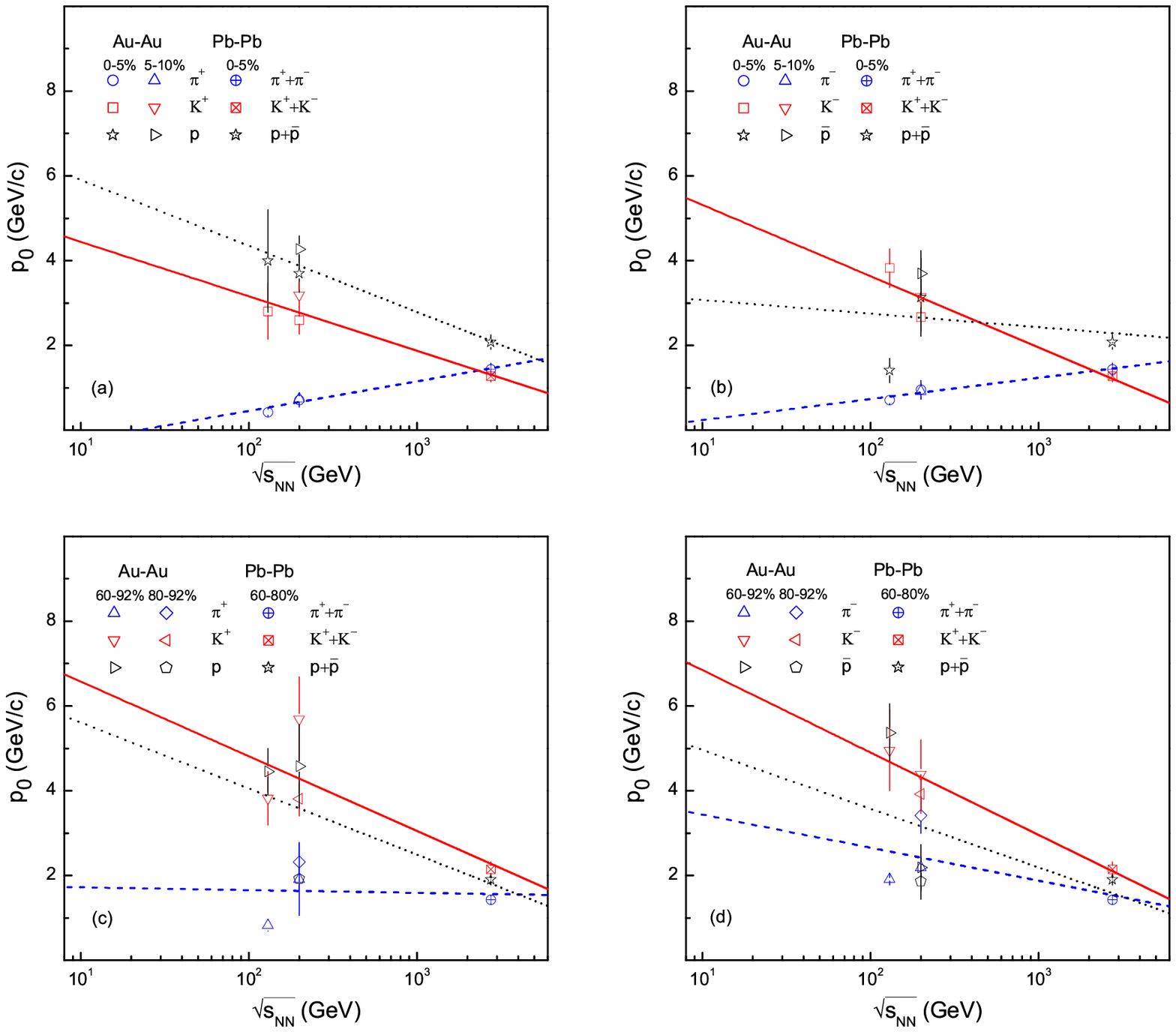}
\end{center}
\vskip0.5cm {\small Fig. 10. Same as Figure 7, but showing the
dependences of $p_0$ on $\sqrt{s_{NN}}$.}
\end{figure*}

\begin{figure*}
\hskip-1.0cm \begin{center}
\includegraphics[width=16.0cm]{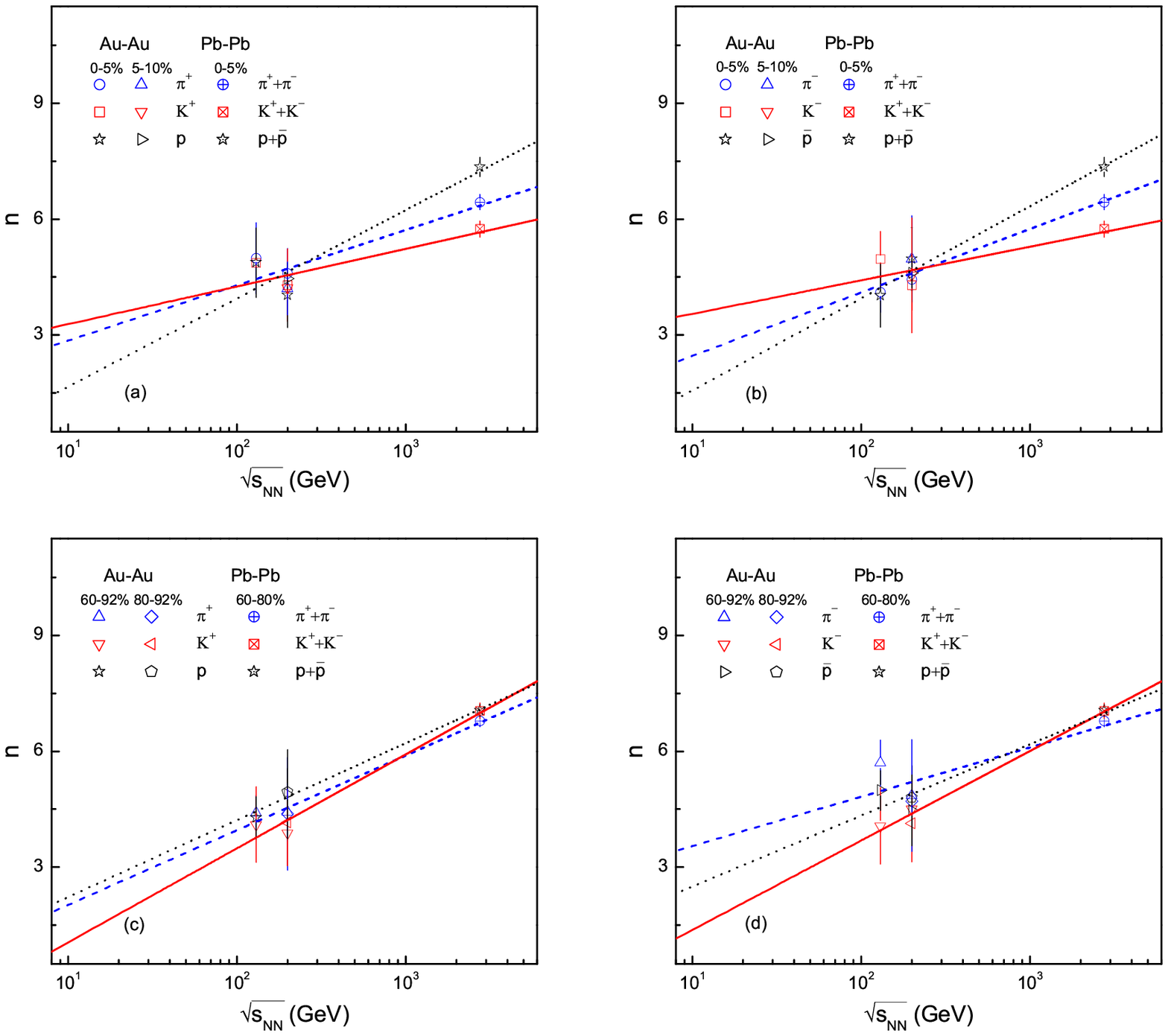}
\end{center}
\vskip0.5cm {\small Fig. 11. Same as Figure 7, but showing the
dependences of $n$ on $\sqrt{s_{NN}}$.}
\end{figure*}

\begin{figure*}
\hskip-1.0cm \begin{center}
\includegraphics[width=16.0cm]{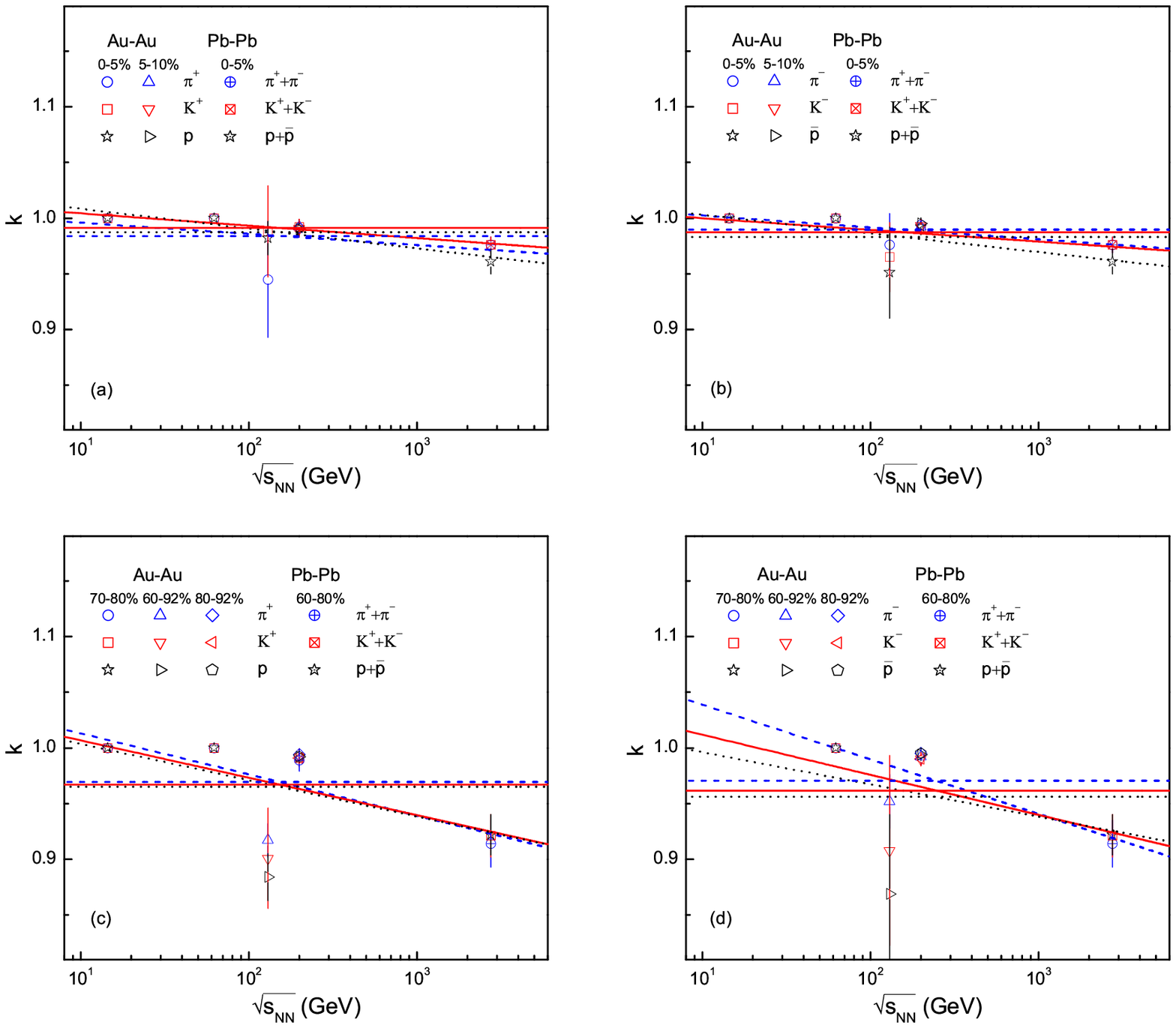}
\end{center}
\vskip0.5cm {\small Fig. 12. Same as Figure 7, but showing the
dependences of $k$ on $\sqrt{s_{NN}}$. The horizontal dashed,
solid, and dotted lines represent the mean values of $k$ over
different energies for charged pions, kaons, and protons
(antiprotons), respectively.}
\end{figure*}

\begin{figure*}
\hskip-1.0cm \begin{center}
\includegraphics[width=16.0cm]{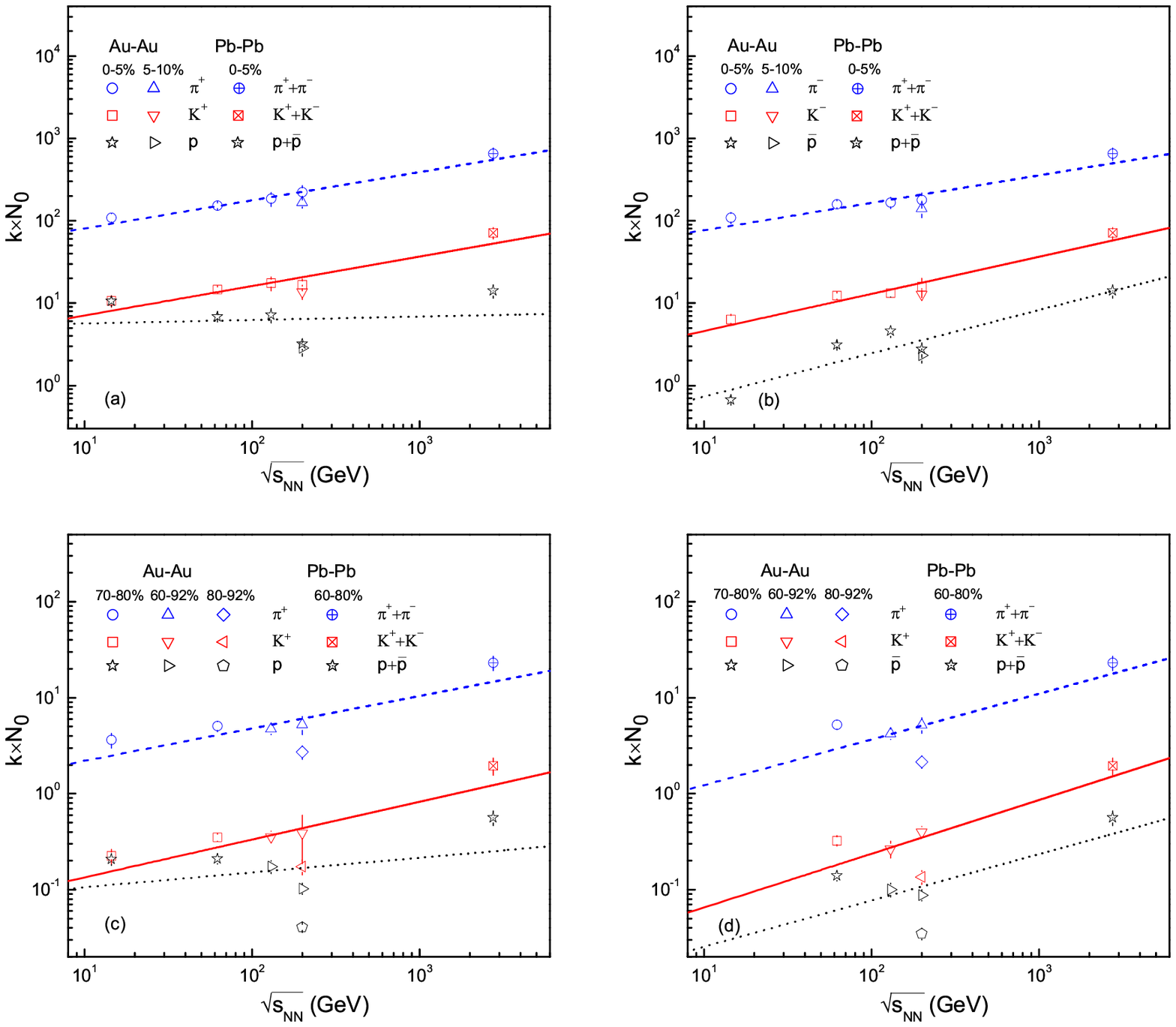}
\end{center}
\vskip0.5cm {\small Fig. 13. Same as Figure 7, but showing the
dependences of $kN_0$ on $\sqrt{s_{NN}}$, where the product $kN_0$
represents the yield of soft excitation process.}
\end{figure*}

\begin{figure*}
\hskip-1.0cm \begin{center}
\includegraphics[width=16.0cm]{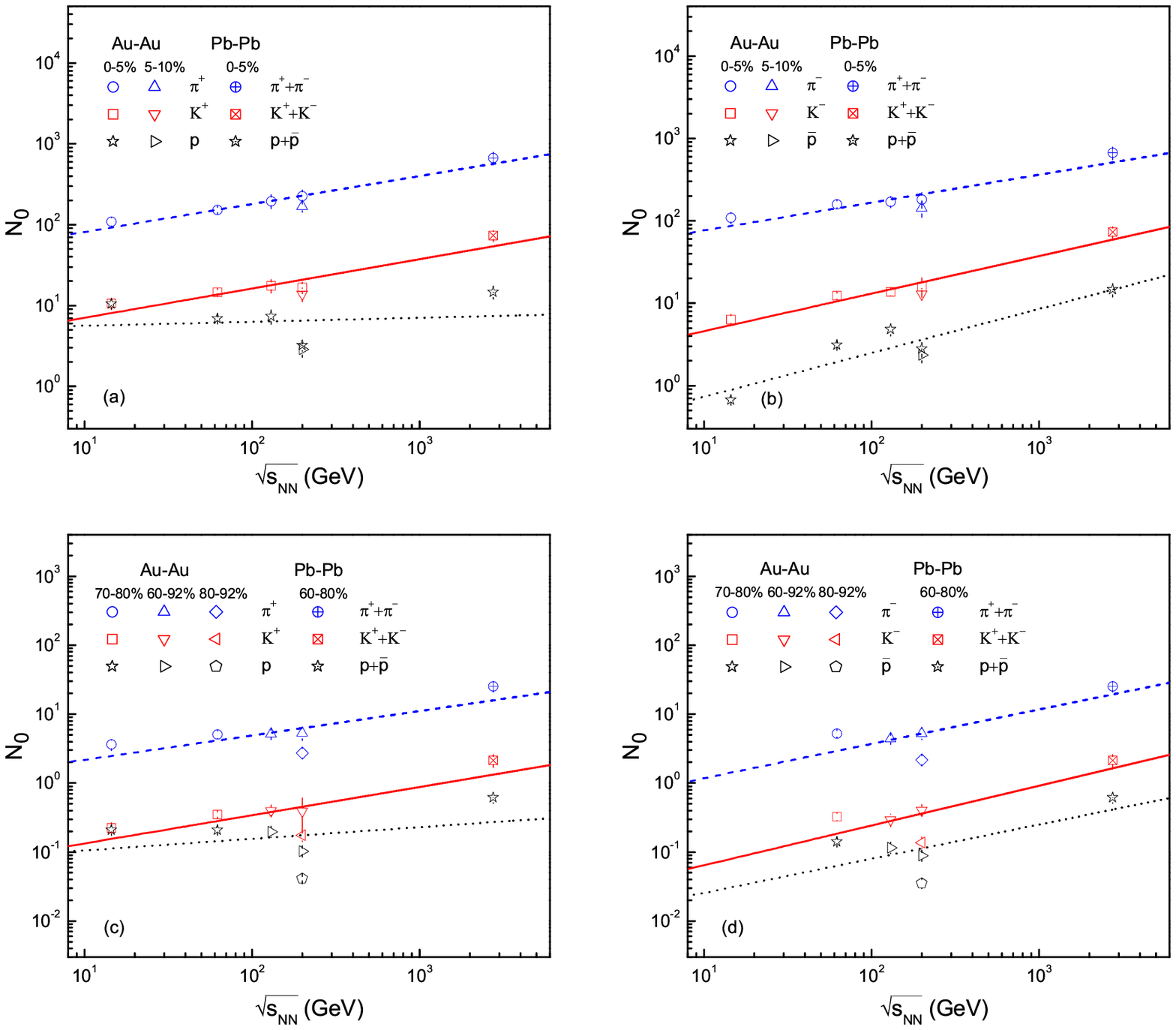}
\end{center}
\vskip0.5cm {\small Fig. 14. Same as Figure 7, but showing the
dependences of $N_0$ on $\sqrt{s_{NN}}$.}
\end{figure*}

\begin{table*}
{\small Table 4. Values of parameters ($a$ and $b$) and
$\chi^2$/dof corresponding to the curves in Figures 8--14. The
function for the curves in Figures 8--12 is
$Y=a+b\ln(\sqrt{s_{NN}})$, where $Y=q$, $\beta_T$, $p_0$, $n$, or
$k$. The function for the curves in Figures 13 and 14 is
$Y=\exp[a+b\ln(\sqrt{s_{NN}})]$, where $Y=kN_0$ or $N_0$. \tiny
\begin{center}
\begin{tabular}{cccccc}
\hline\hline Figure & $Y$ & Main particle & $a$ & $b$ & $\chi^2$/dof \\
\hline
8(a)  & $q$     & $\pi^+$  & $1.053\pm0.039$   & $-0.005\pm0.007$ & $9.284$ \\
      &         & $K^+$    & $0.995\pm0.022$   & $0.008\pm0.004$  & $46.487$ \\
      &         & $p$      & $0.969\pm0.042$   & $0.016\pm0.008$  & $6.489$ \\
8(b)  &         & $\pi^-$  & $1.041\pm0.035$   & $-0.002\pm0.007$ & $6.117$ \\
      &         & $K^-$    & $0.997\pm0.028$   & $0.008\pm0.005$  & $66.313$ \\
      &         & $\bar p$ & $0.968\pm0.042$   & $0.015\pm0.008$  & $200.674$ \\
8(c)  &         & $\pi^+$  & $1.015\pm0.031$   & $0.002\pm0.006$  & $6.725$ \\
      &         & $K^+$    & $0.970\pm0.009$   & $0.010\pm0.002$  & $38.944$ \\
      &         & $p$      & $0.993\pm0.014$   & $0.004\pm0.003$  & $22.203$ \\
8(d)  &         & $\pi^-$  & $1.028\pm0.010$   & $-0.002\pm0.002$ & $0.243$ \\
      &         & $K^-$    & $0.956\pm0.013$   & $0.012\pm0.002$  & $8.679$ \\
      &         & $\bar p$ & $0.992\pm0.017$   & $0.004\pm0.003$  & $15.919$ \\
\hline
9(a)  &$\beta_T$& $\pi^+$  & $0.591\pm0.038$   & $0.003\pm0.007$  & $1.085$ \\
      &         & $K^+$    & $0.556\pm0.048$   & $0.008\pm0.009$  & $2.261$ \\
      &         & $p$      & $0.515\pm0.052$   & $0.015\pm0.010$  & $3.215$ \\
9(b)  &         & $\pi^-$  & $0.586\pm0.021$   & $0.002\pm0.004$  & $0.390$ \\
      &         & $K^-$    & $0.544\pm0.020$   & $0.009\pm0.004$  & $0.318$ \\
      &         & $\bar p$ & $0.516\pm0.026$   & $0.013\pm0.005$  & $1.450$ \\
9(c)  &         & $\pi^+$  & $0.546\pm0.070$   & $0.007\pm0.013$  & $2.362$ \\
      &         & $K^+$    & $0.480\pm0.055$   & $0.016\pm0.010$  & $1.000$ \\
      &         & $p$      & $0.413\pm0.040$   & $0.022\pm0.008$  & $1.623$ \\
9(d)  &         & $\pi^-$  & $0.541\pm0.016$   & $0.011\pm0.003$  & $0.065$ \\
      &         & $K^-$    & $0.495\pm0.033$   & $0.015\pm0.006$  & $0.328$ \\
      &         & $\bar p$ & $0.340\pm0.049$   & $0.034\pm0.009$  & $1.496$ \\
\hline
10(a) &$p_{0}$  & $\pi^+$  & $-0.941\pm0.202$  & $0.303\pm0.034$  & $1.594$ \\
      &         & $K^+$    & $5.731\pm0.714$   & $-0.558\pm0.119$ & $1.245$ \\
      &         & $p$      & $7.462\pm0.669$   & $-0.676\pm0.112$ & $0.874$ \\
10(b) &         & $\pi^-$  & $-0.266\pm0.165$  & $0.217\pm0.028$  & $0.616$ \\
      &         & $K^-$    & $6.999\pm0.854$   & $-0.731\pm0.143$ & $1.437$ \\
      &         & $\bar p$ & $3.392\pm 2.492$  & $-0.139\pm0.417$ & $12.848$ \\
10(c) &         & $\pi^+$  & $1.784\pm1.589$   & $-0.028\pm0.266$ & $20.942$ \\
      &         & $K^+$    & $8.341\pm2.403$   & $-0.766\pm0.402$ & $2.815$ \\
      &         & $p$      & $7.168\pm2.886$   & $-0.677\pm0.483$ & $24.165$ \\
10(d) &         & $\pi^-$  & $4.211\pm1.739$   & $-0.338\pm0.291$ & $110.619$ \\
      &         & $K^-$    & $8.794\pm0.695$   & $-0.845\pm0.116$ & $0.451$ \\
      &         & $\bar p$ & $6.347\pm3.638$   & $-0.602\pm0.609$ & $35.699$ \\
\hline
11(a) & $n$     & $\pi^+$  & $1.426\pm1.084$   & $0.621\pm0.182$  & $0.582$ \\
      &         & $K^+$    & $2.307\pm0.940$   & $0.423\pm0.157$  & $51.514$ \\
      &         & $p$      & $-0.620\pm1.305$  & $0.992\pm0.219$  & $6.059$ \\
11(b) &         & $\pi^-$  & $0.819\pm0.640$   & $0.713\pm0.107$  & $0.172$ \\
      &         & $K^-$    & $2.677\pm0.874$   & $0.378\pm0.146$  & $0.458$ \\
      &         & $\bar p$ & $-0.815\pm 0.530$ & $1.035\pm0.089$  & $0.123$ \\
11(c) &         & $\pi^+$  & $0.091\pm0.432$   & $0.839\pm0.072$  & $0.188$ \\
      &         & $K^+$    & $-1.375\pm0.693$  & $2.430\pm0.267$  & $0.185$ \\
      &         & $p$      & $0.242\pm0.315$   & $0.864\pm0.053$  & $0.087$ \\
11(d) &         & $\pi^-$  & $2.273\pm1.363$   & $0.554\pm0.228$  & $1.451$ \\
      &         & $K^-$    & $-0.941\pm0.429$  & $1.006\pm0.072$  & $0.056$ \\
      &         & $\bar p$ & $0.660\pm0.877$   & $0.799\pm0.147$  & $0.598$ \\
\hline
12(a) & $k$     & $\pi^+$  & $1.006\pm0.027$   & $-0.004\pm0.005$ & $12.283$ \\
      &         & $K^+$    & $1.015\pm0.004$   & $-0.005\pm0.001$ & $6.715$ \\
      &         & $p$      & $1.026\pm0.008$   & $-0.008\pm0.002$ & $4.428$ \\
12(b) &         & $\pi^-$  & $1.013\pm0.010$   & $-0.005\pm0.002$ & $2.861$ \\
      &         & $K^-$    & $1.010\pm0.016$   & $-0.005\pm0.003$ & $5.422$ \\
      &         & $\bar p$ & $1.020\pm0.024$   & $-0.007\pm0.005$ & $10.366$ \\
12(c) &         & $\pi^+$  & $1.050\pm0.043$   & $-0.016\pm0.008$ & $27.256$ \\
      &         & $K^+$    & $1.040\pm0.049$   & $-0.015\pm0.009$ & $16.334$ \\
      &         & $p$      & $1.036\pm0.059$   & $-0.014\pm0.011$ & $30.668$ \\
12(d) &         & $\pi^-$  & $1.088\pm0.041$   & $-0.021\pm0.007$ & $10.944$ \\
      &         & $K^-$    & $1.048\pm0.074$   & $-0.016\pm0.013$ & $34.967$ \\
      &         & $\bar p$ & $1.025\pm0.109$   & $-0.013\pm0.019$ & $54.566$ \\
\hline
13(a) & $kN_0$  & $\pi^+$  & $3.603\pm0.235$   & $0.342\pm0.045$  & $2.604$ \\
      &         & $K^+$    & $1.132\pm0.378$   & $0.358\pm0.072$  & $3.700$ \\
      &         & $p$      & $1.638\pm0.867$   & $0.042\pm0.164$  & $27.912$ \\
13(b) &         & $\pi^-$  & $3.569\pm0.344$   & $0.334\pm0.065$  & $2.535$ \\
      &         & $K^-$    & $0.484\pm0.286$   & $0.451\pm0.054$  & $3.152$ \\
      &         & $\bar p$ & $-1.520\pm0.542$  & $0.525\pm0.103$  & $18.802$ \\
13(c) &         & $\pi^+$  & $0.010\pm0.630$   & $0.338\pm0.119$  & $16.824$ \\
      &         & $K^+$    & $-2.915\pm0.690$  & $0.394\pm0.131$  & $20.334$ \\
      &         & $p$      & $-2.596\pm1.136$  & $0.153\pm0.215$  & $473.659$ \\
13(d) &         & $\pi^-$  & $-0.892\pm1.057$  & $0.476\pm0.187$  & $34.078$ \\
      &         & $K^-$    & $-4.022\pm1.130$  & $0.560\pm0.200$  & $36.881$ \\
      &         & $\bar p$ & $-4.783\pm1.450$  & $0.482\pm0.257$  & $83.614$ \\
\hline
14(a) & $N_0$   & $\pi^+$  & $3.597\pm0.235$   & $0.346\pm0.045$  & $2.693$ \\
      &         & $K^+$    & $1.117\pm0.379$   & $0.363\pm0.072$  & $3.717$ \\
      &         & $p$      & $1.612\pm0.873$   & $0.049\pm0.166$  & $26.466$ \\
14(b) &         & $\pi^-$  & $3.556\pm0.345$   & $0.338\pm0.065$  & $2.522$ \\
      &         & $K^-$    & $0.474\pm0.284$   & $0.455\pm0.054$  & $3.052$ \\
      &         & $\bar p$ & $-1.540\pm0.556$  & $0.533\pm0.106$  & $18.784$ \\
14(c) &         & $\pi^+$  & $-0.042\pm0.647$  & $0.354\pm0.123$  & $18.303$ \\
      &         & $K^+$    & $-2.957\pm0.705$  & $0.409\pm0.134$  & $21.883$ \\
      &         & $p$      & $-2.633\pm1.158$  & $0.168\pm0.220$  & $517.425$ \\
14(d) &         & $\pi^-$  & $-0.984\pm1.076$  & $0.498\pm0.190$  & $36.226$ \\
      &         & $K^-$    & $-4.071\pm1.141$  & $0.576\pm0.202$  & $39.466$ \\
      &         & $\bar p$ & $-4.807\pm1.484$  & $0.495\pm0.263$  & $93.005$ \\
\hline
\end{tabular}
\end{center}}
\end{table*}

With increasing $\sqrt{s_{NN}}$, $p_0$ decreases in some cases and
$n$ increases obviously, their function $f_H(p_T)$ describes a
wider $p_T$ range, though their tendencies seems to be opposite to
each other. According to the relation $n=1/(q-1)$ [14], we know
that the hard scattering process corresponds to a larger $q$ which
makes the collisions be farther away from the equilibrium state
when comparing with the soft excitation process. According to the
relation $p_0=T_0/(q-1)=T_0n$ [14], we know that the hard process
corresponds to a higher $T_0$, which results in more violent
collisions than is the soft process. This observation is a natural
result. The differences in $p_0$ and in $n$ for central and
peripheral collisions, as well as for different particle
productions, are not obvious. This reflects the fact that, in the
hard process, the participant valence quarks collide deeply at the
initial state where the spectator nucleons have no effect.

The relative contribution ($k$) of the soft excitation process
decreases slightly and the relative contribution ($1-k$) of the
hard scattering process increases slightly with the increase of
$\sqrt{s_{NN}}$. This is consistent with the theoretical
expectation [34, 35], and inconsistent with the extraction based
on the numbers of participating nucleons and binary
nucleon-nucleon collisions [36, 37]. In our opinion, at higher
energy, both the participant valence quarks have more probability
to approach each other and to perform interactions. Then, the hard
process has more contribution to the $p_T$ spectrum. However, the
participant gluons and/or sea quarks have less time to perform
interactions due to the higher pass speed at higher energy, which
means that the soft process contributes less to the $p_T$
spectrum. The differences in $k$ for central and peripheral
collisions, as well as for different particle productions, are not
obvious. This also reflects the fact that in the hard process the
participant valence quarks collide deeply at the initial state
where the spectator nucleons have no effect.

Generally, the yields of the soft process and the soft plus hard
processes shown in Figures 13 and 14 increase with the increase of
$\sqrt{s_{NN}}$ except for protons that saturate due to the
limited proton numbers in the participant nuclei. Obviously, the
yield for $\pi^{\pm}$ is greater than that for $K^{\pm}$, and much
greater than that for $\bar p$. The yield for central collisions
is greater than that for peripheral collisions. The yield for
positive mesons is slightly greater than that for negative mesons.
These tendencies of the yields are natural results due to the
experimental data analyzed in the present work.

From the above discussions, in particular from Figures 7--14 and
Tables 1--4 one can see that, not only for the dependence on
$\sqrt{s_{NN}}$ but also for the dependence on centrality, all the
free parameters in the improved Tsallis distribution and in the
inverse power-law seem to be independent of isospin. This means
that electromagnetic interactions play a minor role in both the
soft and hard processes.

We would like to point out that, as can be seen from Tables 1 and
2, the model fitting is not very good in a few cases, because the
values of $\chi^2$/dof are very large for these cases. That does
not mean that the model cannot describe the particle distributions
in heavy ion collisions. In fact, in most cases, we have obtained
appropriate values of $\chi^2$/dof which imply that the model
works well. The very large values of $\chi^2$/dof are obtained due
to abnormally small errors. In the case of using a relative error
being 5\%, the values of $\chi^2$/dof are appropriate. From Tables
3 and 4 one can see that the values of $\chi^2$/dof are very large
in many cases. That means that the relations
$Y=a+b\ln(\sqrt{s_{NN}})$ and $Y=\exp[a+b\ln(\sqrt{s_{NN}})]$
assumed by us do not work well.

In the above discussions, we have used six free parameters, the
kinetic freeze-out temperature $T_0$, the entropy index $q$, the
radial velocity flow $\beta_T$, the fraction $k$ of soft
component, $p_0$, and $n$. The meanings of the first four
parameters are clear. The meanings of the last two parameters can
result from the relations $p_0=T_0/(q-1)$ and $n=1/(q-1)$ [14]
which show that $p_0$ has a similar meaning to $T_0$ and $n$ has
an opposite meaning in comparison with $q$. In many cases, the
mass-dependent tendencies of these parameters are not obvious
because they are just from the model fittings. In addition, to
obtain the mass-dependent tendencies of these parameters, we need
more types of particles.

The present work shows that the interacting systems at the LHC
have a higher excitation and larger expansion than those at the
RHIC due to a greater energy depositions at the LHC. The central
collisions have a higher excitation and larger expansion than the
peripheral collisions due to greater energy depositions in the
central collisions. Both the central and peripheral collisions are
approximately in the equilibrium states, though the peripheral
collisions are closer to the equilibrium state. Comparing with
that at the RHIC, the transverse momentum spectrum at the LHC has
a lower fraction of soft component, because the participant gluons
and/or sea quarks have less time to perform interactions in the
case of the heavy ions having higher pass speed at the LHC.

It should be noted that the same information can be obtained from
other methods, and the values for the same quantities from
different methods are different. In other words, the results are
model dependent. In particular, for the kinetic freeze-out
temperature or the radial flow velocity, different methods can be
regarded as different `thermometers' or `speedometers'. To make a
comparison for the results obtained from different methods, we
need to structure a standard method which can be used to make
comparison with others. Or, we can use the alternative method in
which $T_0$ is regarded as the intercept in the linear relation
between $T$ and $m_0$ [11, 18--20], and $\beta_T$ is regarded as
the slope in the linear relation between $\langle p_T \rangle$ and
$\langle m \rangle$ [15--17].
\\

{\section{Conclusions}}

We summarize here our main observations and conclusions.

(a) The $p_T$ spectra of $\pi^{\pm}$, $K^{\pm}$, $p$, and $\bar p$
produced in Au-Au and Pb-Pb collisions over an energy
$\sqrt{s_{NN}}$ range from 14.5 GeV to 2.76 TeV have been
analyzed. For the spectra with a narrow $p_T$ range, the improved
Tsallis distribution which describes the soft process is used. For
the spectra with a wide $p_T$ range, the superposition of the
improved Tsallis distribution and the inverse power-law which
describes the hard process is used. The modelling results are in
approximate agreement with the experimental data measured by the
STAR, PHENIX, and ALICE Collaborations. Some parameters are
extracted due to the fittings.

(b) Both the extracted $T_0$ and $\beta_T$ increase with the
increase of $\sqrt{s_{NN}}$, which indicates a higher excitation
and larger expansion of the interesting system at the LHC. Both
the values of $T_0$ and $\beta_T$ in central collisions are
slightly larger than those in peripheral collisions. The slight
differences in $T_0$ for different particles are observed in some
cases. This confirms the mass-dependent differential kinetic
freeze-out scenario. An evidence of mass-dependent $\beta_T$ is
observed in most cases. A heavy particle corresponds to a small
$\beta_T$ due to its large inertia. The differences in $\beta_T$
for different particles decrease with the increase of
$\sqrt{s_{NN}}$. The mass-dependent effect of $\beta_T$ can be
neglected in a strong flow field at the LHC.

(c) The parameter $q$ increases slightly with the increase of
$\sqrt{s_{NN}}$, but the dependence of $q$ on $\sqrt{s_{NN}}$ is
not obvious. The parameters $q$ in central and peripheral
collisions are very small, which means that the two types of
collisions are in the nearly equilibrium state respectively,
though a slight larger $q$ seems to be observed in central
collisions. In most cases, the differences in $q$ for different
particles are not obvious. The relative contribution $k$ of the
soft process decreases slightly with the increase of
$\sqrt{s_{NN}}$. At the LHC, the participant gluons and/or sea
quarks have less time to perform interactions due to higher pass
speed, which means that the soft process contributes less to the
$p_T$ spectrum. The differences in $k$ for central and peripheral
collisions, as well as for different particle productions, are not
obvious.

(d) With increasing $\sqrt{s_{NN}}$, $p_0$ decreases in some cases
and $n$ increases obviously, their function $f_H(p_T)$ describes a
wider $p_T$ range. The differences in $p_0$ and in $n$ for central
and peripheral collisions, as well as for different particle
productions, are not obvious. This reflects the fact that, in the
hard process, the participant valence quarks collide deeply at the
initial state where the spectator nucleons have no effect. The
relative contribution $1-k$ of the hard process increases slightly
with the increase of $\sqrt{s_{NN}}$. At higher energy, both the
participant valence quarks have more probability to approach each
other and to perform interactions. Then, the hard process has more
contribution to the $p_T$ spectrum at the LHC. Our conclusion is
in agreement with the theoretical prediction based on QCD.

(e) The yields of the soft process and the soft plus hard
processes increase with the increase of $\sqrt{s_{NN}}$ except for
protons that saturate due to the limited proton numbers in the
participant nuclei. The yield for $\pi^{\pm}$ is greater than that
for $K^{\pm}$, and much greater than that for $\bar p$. The yield
for central collisions is greater than that for peripheral
collisions. The yield for positive mesons is slightly greater than
that for negative mesons. These tendencies of the yields appear
due to the experimental data themselves. In fact, the yields are
the normalization constants. Not only for the dependence on
$\sqrt{s_{NN}}$ but also for the dependence on centrality, all the
free parameters in the improved Tsallis distribution and in the
inverse power-law seem to be independent of isospin. This means
that electromagnetic interactions play a minor role in both the
soft and hard processes.
\\

{\bf Conflict of Interests}

The authors declare that there is no conflict of interests
regarding the publication of this paper.
\\

{\bf Acknowledgments}

This work was supported by the National Natural Science Foundation
of China under Grant No. 11575103 and the US DOE under contract
DE-FG02-87ER40331.A008.
\\

\end{multicols}
\end{document}